\documentclass[aps,pra,twocolumn,reprint,showkeys,preprintnumbers,amsmath,amssymb,floatfix,a4paper,longbibliography,superscriptaddress]{revtex4-2}

\usepackage{amsmath}
\usepackage{amsfonts}
\usepackage{amssymb}
\usepackage{bm}
\usepackage{bbm}
\usepackage{color}
\usepackage{graphicx}

 \usepackage{hyperref}
\hypersetup{breaklinks=true}
\usepackage{dcolumn}
\usepackage{array}

\usepackage{newtxtext,newtxmath} 

\usepackage{orcidlink}

\usepackage[T1]{fontenc}   

\newcommand{\figwidth}{1\columnwidth}
\newcommand{\nqubit}{K_{\rm q}}
\newcommand{\nqubittot}{K_{\rm q}^{\rm tot}}
\newcommand{\Ntrotter}{N_{\rm ST}}
\newcommand{\Nsites}{N}
\newcommand{\Nshots}{N_{\text{shot}}}
\newcommand{\Ntwoq}{N_{\text{2q}}}

\newcommand{\spin}{S}
\newcommand{\Ms}{M}
\newcommand{\dt}{\Delta t}
\newcommand{\dtau}{\Delta \tau}
\newcommand{\Msi}[1]{\Ms_{#1}}
\newcommand{\Sop}[2]{\hat{S}_{#2}^{#1}}
\newcommand{\boldSop}[1]{\hat{\bm{S}}_{#1}}
\DeclareMathOperator{\sym}{sym}
\DeclareMathOperator{\bin}{bin}
\DeclareMathOperator{\trace}{tr}
\newcommand{\ketq}[1]{|{#1}\rangle_{\rm q}}
\newcommand{\braq}[1]{{}_{\rm q}\langle{#1}|}
\newcommand{\UDicke}{U_{\rm Dicke}}
\newcommand{\UST}{U_{\rm ST}}
\newcommand{\NHq}{L}
\newcommand{\NHqcompact}{L_{\rm compact}}
\newcommand{\avNHqcompact}{\overline{L}_{\rm compact}}
\newcommand{\Hamiltonian}{\mathcal{H}}
\newcommand{\kmin}{\kappa_0}
\newcommand{\kmax}{\kappa_1}
\newcommand{\quditd}{d}
\newcommand{\GellMann}{\lambda}
\newcommand{\erroroneq}{\epsilon_{\rm 1q}}
\newcommand{\errortwoq}{\epsilon_{\rm 2q}}
\newcommand{\Avdisc}{\overline{\Delta}}
\newcommand{\W}{W}	
\newcommand{\SMone}{1_{\spin-\Ms}}
\newcommand{\bigO}{\mathcal{O}}
\begin{document}

\title{Comparison of encoding schemes for  quantum computing of \boldmath{$\spin > 1/2$} spin chains}
\date{\today}
\author{Erik L\"otstedt\,\orcidlink{0000-0003-2555-8660}}
\email{loetstedte@riken.jp}
\affiliation{Department of Chemistry, School of Science, The University of Tokyo, 7-3-1 Hongo, Bunkyo-ku, Tokyo 113-0033, Japan}
\affiliation{Trapped Ion Quantum Computer Team, TRIP Headquarters, RIKEN,   2-1 Hirosawa, Wako, Saitama 351-0198, Japan}
\altaffiliation{Presently also affiliated with (a) RIKEN Center for Interdisciplinary Theoretical and Mathematical Sciences (iTHEMS) and (b) the Computational Condensed Matter Physics Laboratory, RIKEN Cluster for Pioneering Research}
\author{Kaoru Yamanouchi\,\orcidlink{0000-0003-3843-988X}}
\affiliation{Department of Chemistry, School of Science, The University of Tokyo, 7-3-1 Hongo, Bunkyo-ku, Tokyo 113-0033, Japan}
\affiliation{Trapped Ion Quantum Computer Team, TRIP Headquarters, RIKEN,   2-1 Hirosawa, Wako, Saitama 351-0198, Japan}
\affiliation{Institute for Attosecond Laser Facility, The University of Tokyo, 7-3-1 Hongo, Bunkyo-ku, Tokyo 113-0033, Japan}

\begin{abstract}
We compare four different encoding schemes  for  the quantum computing of spin chains with a spin quantum number  $\spin>1/2$: a compact mapping, a direct (or one-hot) mapping, a Dicke mapping, and a qudit mapping. The three different qubit encoding schemes are assessed by  conducting Hamiltonian simulation  for $1/2\le\spin\le5/2$ using a trapped-ion quantum computer. The qudit mapping is tested by running simulations with a simple noise model.
The Dicke mapping, in which the  spin states are encoded as superpositions of multi-qubit states,  is found to be the most efficient because of the small number of terms in the qubit Hamiltonian. We also investigate the $\spin$-dependence  of the time step length $\dtau$ in the Suzuki-Trotter approximation and find that 
  $\dtau$  should be inversely proportional to $\spin$ in order to obtain the same accuracy for all  $\spin$.
\end{abstract}


\maketitle
%
\newpage 

\section{Introduction}\label{Sec:Intro}
One of the main applications of quantum computers is the simulation of  quantum systems such as molecules \cite{Caoetal_review2019,Baueretal2020,McArdleetal_review2020,Claudino2022,Pyrkov2023,Mazzola2024,Weidman2024} and solid-state crystals  \cite{Baueretal2020,Alexeev2024}. While fault tolerance may be required for quantum computers to be truly useful in quantum simulation \cite{Katabarwa2024}, pioneering calculations using noisy intermediate-scale quantum (NISQ \cite{Preskill2018}) devices have been carried out  for small molecules such as 
H$_2$O \cite{Eddinsetal2021}, 
CH$_4$\cite{Khan2023},
C$_4$H$_6$ \cite{OBrien2023}, 
Li$_2$O \citep{ZhaoGoings2023}, and F$_2$ \cite{Guo2024}. However,  because of the  inherent noise of NISQ devices and because of the deep circuits for the wave function ansatz,  the number of qubits in quantum chemistry calculations has been limited to around 10 as in the recent calculation of the potential energy curve of F$_2$ by Guo \emph{et al.}\ \cite{Guo2024}, in which 12 qubits were used.

On the other hand, interacting 1/2 spins on a lattice or in a molecule is a type of  quantum system which is well suited for simulation on quantum computers. This is because the two quantum states  $|{\uparrow}\rangle$ and $|{\downarrow}\rangle$ of a spin-1/2 site  can be straightforwardly mapped to the two qubit states $|0\rangle$ and $|1\rangle$.  The simulation of the quantum dynamics of interacting spins can be used to understand the behavior of magnetic materials at low temperatures. The simplest model is the transverse field Ising model \cite{Gennes1963,Stinchcombe1973}, which has been a popular model for simulations on quantum computers. Successful demonstrations \cite{LammLawrence2018,Bassmanetal2020,Sopena2021,Zhukov2018,Kim2023}
have been carried out using up to 127 qubits \cite{Kim2023a}. A slightly more sophisticated model is the Heisenberg model \cite{Heisenberg1928,VanVleck1945,Pires2021}, different variants of which have been implemented and simulated on quantum computers 
\cite{Zhukov2018,Chiesa2019,SmithKimPollmannKnolle2019,ZhuJohrietal2021,Urbaneketal2021,Bosse2022,%
Kattemoelle2022,Oftelie2022,Tazhigulov2022,LotstedtWangetal2023,LotstedtYamanouchi2024_copy,Rosenberg2024}. In recent simulations,  
 up to 100 qubits were employed  \cite{Yu2023,Chowdhury2024}.

While the Ising and Heisenberg models usually involve 1/2-spins, systems of interacting spins having a spin quantum number $\spin>1/2$ are also interesting. Systems of spins with a large $\spin$ can be realized experimentally in  cluster complexes
\cite{Wang2007,Sharma2014,Baniodeh2018}, 
nanomagnets \cite{Thomas1996,Wernsdorfer1999,Jankowski2021,Zhao2022}, and 
single-molecule magnets \cite{Sessoli1993,Mereacre2007,Liu2014,ZabalaLekuona2021,Nehrkorn2021,Wang2023}. For example, in  \cite{Baniodeh2018}, a ring-like compound including 10 Fe$^{\rm III}$ ($\spin=5/2$) and 10 Gd$^{\rm III}$ ($\spin=7/2$) ions was synthesized. At low temperatures below a few K, the large $\spin$ spin chains realized in molecules like the one reported in  \cite{Baniodeh2018} need to be described by a quantum mechanical model of Heisenberg type.

Spin chains with $\spin>1/2$ are also of theoretical interest 
for high-spin Kitaev models \cite{Baskaran2008,Stavropoulos2019,Pohle2024}
as well as for the  ``Haldane gap'' \cite{Haldane1983,NightingaleBlote1986,Affleck1989,Jolicoeur2019,Tasaki2025}, which refers to the finite energy difference between the ground state and the first excited state in the one-dimensional, infinitely long Heisenberg spin chain with integer values of $\spin$.  

Even though most of the previous studies on quantum computing of interacting spin systems have concentrated on $\spin=1/2$ systems, there exist a few reports on $\spin=1$ lattices.
Quantum annealing has been considered to find the ground state of $\spin=1$ systems \cite{Cao2021}. 
The real-time dynamics of a single-site $\spin=1$ system was 
considered in \cite{Gnatenko2023,Kuzmak2023} and that of a chain of 31 $\spin=1$ sites described by a Kogut-Susskind Hamiltonian was simulated in \cite{Hayata2025}. In these studies  \cite{Gnatenko2023,Kuzmak2023,Hayata2025},
IBM superconducting quantum computers \cite{ibmq} were employed.  Spin chains having $\spin=1$  with four \cite{Senko2015} and five \cite{Edmunds2024} sites were simulated using trapped 
 ions of ${}^{171}$Yb$^+$  \cite{Senko2015} and ${}^{40}$Ca$^+$ \cite{Edmunds2024}.

The number of spin states 
at each lattice site is $2\spin +1$, and the total number of spin states for a spin lattice with $\Nsites$ sites is equal to
 $(2\spin+1)^{\Nsites}$, 
which increases exponentially with increasing $\Nsites$. It is in general a difficult task to simulate the spin dynamics of 
spin lattices using classical computers \cite{Wu2024} even though sophisticated Monte Carlo methods \cite{Kawashima1994,Todo2001}, density matrix renormalization group methods \cite{WhiteHuse1993,Ueda2011,Yu2021,Hagymasi2022}, numerical diagonalization techniques \cite{Nakano2019,Nakano2022,Sattler2024}, and semiclassical methods \cite{Remund2022} have been developed.
Therefore, it is worthwhile to develop efficient  algorithms for spin lattices with $\spin>1/2$ to be simulated on quantum computers, because the number of spin states which can be stored in memory in qubit-based quantum computing scales in principle exponentially with the number of qubits.

In the present study, we investigate  how spin chain dynamics  with arbitrary spin $\spin$ can be simulated using quantum computers. Because each spin site has $2\spin +1$ states, which is larger than two when $\spin>1/2$, more than one qubit must be used for each site. We investigate three kinds of mappings of the spin states to the qubit states: (i) a compact mapping, (ii) a direct mapping, and (iii) a mapping developed for spin-chain simulations using classical computers \cite{Kawashima1994,Todo2001}, which we refer to as the Dicke mapping. The Dicke mapping was recently suggested as a suitable mapping for the simulation of $\spin>1/2$ lattices using neutral atoms \cite{Maskara2025}. In addition, in view of the recent proposal \cite{Edmunds2024} of using a qudit quantum computer for the simulation of  an $\spin=1$ chain, we also consider (iv) a qudit mapping where each spin-$\spin$ lattice site is mapped to a qudit having $\quditd = 2\spin+1$ levels.
 We derive the qubit and qudit Hamiltonians for these four different encoding schemes and compare their performances 
 by conducting simulations of a two-site spin model for $\spin=1$ and $3/2$. The simulations are performed using the Quantinuum trapped-ion quantum computer H1-1 \cite{QuantinuumH11_2024} in the case of the qubit mappings.
 In the case of the qudit mapping, we conduct the simulations on a classical computer using  a completely depolarizing  error model to approximately incorporate the effect of the noise.
We further study the performance of the Dicke mapping by simulating  a  four-site spin chain with varying $\spin$ in the range  of $1/2\le \spin \le 5/2$. We also investigate the $\spin$-dependence of the time step size $\dtau$ in the Suzuki-Trotter approximation.

\section{Theory}\label{Sec:Theory}

\subsection{Hamiltonian and spin states}
As the model for spin-spin interaction, we consider the simplest form of the Heisenberg Hamiltonian \cite{VanVleck1945}, 
\begin{align}\label{Eq:HeisenbergH}
	\Hamiltonian&=\frac{J}{\hbar^2}\sum_{(m,n)}\boldSop{m} \cdot \boldSop{n} 
	\nonumber
	\\
	&= 
	\frac{J}{\hbar^2}\left(
    \sum_{(m,n)}\Sop{z}{m}\Sop{z}{n} +\frac{1}{2} \sum_{(m,n)}\Sop{+}{m}\Sop{-}{n}+
	\frac{1}{2} \sum_{(m,n)}\Sop{-}{m}\Sop{+}{n}
     \right),
\end{align}
where the sum over $m$ and $n$ is taken over the interacting sites on the lattice, $\boldSop{n}=(\Sop{x}{n},\Sop{y}{n},\Sop{z}{n})$ is the vector of spin operators acting on site $n$, and $\Sop{\pm}{n}=\Sop{x}{n}\pm i\Sop{y}{n}$. 
The coupling constant $J$ can be removed by a scaling of the Hamiltonian so that energy is measured in units of $J$ and time in units of $\hbar/J$.
The lattice spin states are defined as 
\begin{equation}\label{Eq:lattice_spin_states}
	|\vec{\Ms}\rangle=|\Msi{\Nsites-1},\Msi{\Nsites-2},\ldots,\Msi{0}\rangle,
\end{equation}
where $\Nsites$ is the number of sites and $\Msi{n}$ is the magnetic quantum number of site $n$ taking values of 
$\Msi{n}=-\spin,-\spin+1,\ldots,\spin$.
The actions of 
$\Sop{z}{n}$ and $\Sop{\pm}{n}$ on the spin state $|\Msi{n}\rangle$  at a site $n$
are defined as usual as
\begin{equation}\label{Eq:SzM}
	\Sop{z}{n}|\Msi{n}\rangle = \hbar \Msi{n}|\Msi{n}\rangle
\end{equation}
and
\begin{equation}\label{Eq:SpmM}
 \qquad \Sop{\pm}{n}|\Msi{n}\rangle = \hbar\sqrt{\spin(\spin+1)-\Msi{n}(\Msi{n}\pm 1)}|\Msi{n}\pm 1\rangle.
\end{equation}
We consider spin lattices where all the sites have the same value of $\spin$, but the generalization to non-uniform lattices in which the values of $\spin$ can be different at different sites is straightforward. 
\subsection{Qubit mappings}\label{subsec:QubitMappings}
In order to map the spin states \eqref{Eq:lattice_spin_states} onto qubit or qudit states, which can be manipulated on a quantum computer, we consider four different kind of mappings: (i) a compact mapping, (ii) a direct mapping,  (iii) a mapping in terms of composite many-qubit states, referred to as the Dicke mapping, and (iv) a qudit mapping.

(i) \emph{Compact mapping.} In this mapping, which is also referred to as a binary mapping \cite{Sawayaetal2020}, the spin states are mapped to qubit states according to 
\begin{equation}\label{Eq:CompactDef}
|\Ms\rangle = \ketq{\bin(\Ms+\spin)},
\end{equation} 
where $\bin(x)$ denotes the binary representation of $x$, and a subscript q is attached to the qubit state to distinguish it from the spin state.
The compact mapping requires $\nqubit=\lceil\log_2(2\spin+1)\rceil$ qubits per site, where $\lceil\cdot\rceil$ denotes the ceiling function.
For example, for $\spin=1$, we have three states ($\Ms=-1,$ 0, 1), which are mapped to two-qubit states according to
\begin{equation}
|{-1}\rangle = \ketq{00},
\text{ }
|0\rangle = \ketq{01}, 
\text{ }
\text{and  }
|1\rangle = \ketq{10}.
\end{equation}
The $\ketq{11}$ state is not used for $\spin =1$. To describe a two-site lattice, we need four qubits and use the mapping $|{-1},{-1}\rangle =\ketq{0000}$, 
$|{-1},0\rangle =\ketq{0001}$, and so on.

(ii) \emph{Direct mapping.} In the direct mapping, also called unary encoding \cite{Sawayaetal2020} or  one-hot mapping \cite{Hadfield2019}, the spin state $|\Ms\rangle$ is mapped to a qubit state where the $(\Ms+\spin)$th qubit is in the excited $\ketq{1}$ state,
\begin{equation}\label{Eq:DirectDef}
|\Ms\rangle = X_{\Ms+\spin}\ketq{\overline{0}},
\end{equation}
where $\ketq{\overline{0}}=\ketq{0 0 \cdots 0}$ is the all-zero state, and $X_k$ is a Pauli $\sigma_x$ operator acting on qubit $k$. The number of qubits required per site in the direct mapping is 
$\nqubit = 2\spin +1$.

(iii) \emph{Dicke mapping.} In this mapping, we represent the spin states in terms of a symmetric superposition of qubit states,
\begin{equation}\label{Eq:Dickestate_def}
|\Ms\rangle = 
\sym\mathopen{}\left(
\ketq{\SMone} 
               \right)\mathclose{} 
\equiv \ketq{D_{\spin,\Ms}},
\end{equation} 
where $\sym(\ketq{\psi})$ denotes the symmetrization operator which symmetrizes a state  $\ketq{\psi}$ (including the normalization constant) with respect to qubit exchange and $\ketq{\SMone}$ is defined as
\begin{equation}\label{Eq:SmMhotstate}
	\ketq{\SMone}=X_{\spin-\Ms-1}X_{\spin-\Ms-2}\cdots X_{0}\ketq{\overline{0}}.
\end{equation}
In the Dicke mapping, $\nqubit=2\spin$ qubits per site are employed. Equation \eqref{Eq:Dickestate_def} implies that $|\Ms\rangle$ is represented as an equal superposition of the 
$\binom{2\spin}{\spin-\Ms}$ qubit states having 
$\spin-\Ms$ qubits in the excited $\ketq{1}$ state and 
$\spin+\Ms$ qubits in the ground $\ketq{0}$ state. For example, for $\spin=3/2$, we have 
\begin{align}
\left|-\tfrac{3}{2}\right\rangle & = \ketq{111},
\nonumber
\\
\left|-\tfrac{1}{2}\right\rangle &= \frac{1}{\sqrt{3}}\left(\ketq{011}+\ketq{101}+\ketq{110}\right),
\nonumber
\\
\left|\tfrac{1}{2}\right\rangle & = \frac{1}{\sqrt{3}}\left(\ketq{010}+\ketq{001}+\ketq{100}\right),
\text{ and}
\nonumber
\\
\left|\tfrac{3}{2}\right\rangle &=\ketq{000}.
\end{align}
The states $\ketq{D_{\spin,\Ms}}$ defined in Eq.~\eqref{Eq:Dickestate_def} are referred to as Dicke states after R.\ H.\ Dicke, who introduced them  in quantum optics \cite{Dicke1954}.
The Dicke states \eqref{Eq:Dickestate_def} are eigenstates of the  operator
\begin{equation}\label{Eq:S2_qubits}
S^2_{\rm q}=\frac{1}{4}\sum_{k,l=0}^{\nqubit-1}\left(Z_kZ_l +Y_kY_l + X_kX_l\right)
\end{equation}
with the common $\Ms$-independent  eigenvalue  $\spin(\spin+1)$. The sum in Eq.~\eqref{Eq:S2_qubits} is taken over all $\nqubit=2\spin$ qubits, and $X_k$, $Y_k$, $Z_k$ are Pauli operators acting on qubit $k$.  The operator \eqref{Eq:S2_qubits} can be 
viewed as a total spin operator
if each qubit is viewed as a spin-1/2 system with the identification $\ketq{0}=|\tfrac{1}{2}\rangle$ and $\ketq{1}=|{-}\tfrac{1}{2}\rangle$, and the Dicke states \eqref{Eq:Dickestate_def} represent  qubit states having maximal spin $\spin$. 
We note that, as $\spin$ increases, the number of Dicke states ($2\spin+1$) becomes a much smaller fraction of the total number of the qubit states ($2^{2\spin}$).

The approach of describing large $\spin$ states as composite states of spin-1/2 particles  is well known in Monte Carlo simulations of spin lattices \cite{Kawashima1994,Kawashima1995,Todo2001}. 
We note that the Dicke mapping was used in the pioneering simulations of a single-site, $\spin=1$ system in \cite{Gnatenko2023,Kuzmak2023} and was  also used in \cite{Maskara2025} for general $\spin>1/2$. 

Efficient quantum circuits for creating Dicke states have been demonstrated in  \cite{Baertschi2019,Bartschi2022}. The total number of {\sc cnot} gates scales as $\bigO(\nqubit^2)$.
An alternative scheme of preparing Dicke states employing a control Hamiltonian with  Ising-type coupling was proposed in \cite{Stojanovic2023} and a scheme for changing the magnetic quantum number $\Ms$ of a Dicke state was presented in \cite{Stojanovic2022}.

\begin{figure}
\includegraphics[width=\figwidth]{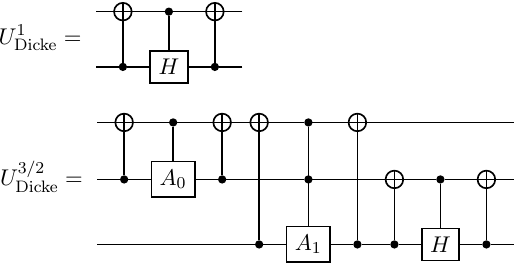}
\caption{\label{Fig1}
Circuit representations of the Dicke unitary operators $\UDicke^{\spin=1}$ and $\UDicke^{\spin=3/2}$.
The Hadamard gate is denoted by $H$.
In the circuit for $\UDicke^{\spin=3/2}$, the  single-qubit $A_0$ gate is defined in matrix form as
$A_0=\left(
\begin{smallmatrix}
	1&-\sqrt{2}\\ 
	\sqrt{2} & 1 
\end{smallmatrix}
\right)/\sqrt{3}$ 
and 
$A_1$ is defined as $A_1=\left(\begin{smallmatrix}\sqrt{2}&-1\\ 1 & \sqrt{2}\end{smallmatrix}\right)/\sqrt{3}$.
The $A_k$ gates are implemented in practice as $R_Y$ gates.
The circuit diagrams in this manuscript are drawn using the {\sc Quantikz} package \cite{Kay2020_Quantikz}.
}
\end{figure}

We define the Dicke circuit $\UDicke^{\spin}$, the $2\spin$-qubit unitary operator implementing the $\sym(\cdot)$ operation in Eq.~\eqref{Eq:Dickestate_def} as
\begin{equation}\label{Eq:UDicke_def}
\UDicke^{\spin} X_{\spin-\Ms-1}X_{\spin-\Ms-2}\cdots X_{0}\ketq{\overline{0}} = \ketq{D_{\spin,\Ms}}.
\end{equation}
We note that $\UDicke^{\spin}$ is independent of $\Ms$. In Fig.~\ref{Fig1}, we show two examples of $\UDicke^{\spin}$ for $\spin=1$ and $\spin=3/2$,  constructed according to the method presented in \cite{Baertschi2019,Bartschi2022}. By direct inspection, we can confirm that the two circuits satisfy Eq.~\eqref{Eq:UDicke_def} for different values of $\Ms$.

(iv) \emph{Qudit mapping.} A qudit is a $\quditd$-level quantum system \cite{Wang2020}, and has been realized as a building block for a quantum computer in superconducting circuits \cite{Blok2021,Nguyen2024}, 
photonic circuits \cite{Chi2022}, and trapped ions \cite{Ringbauer2022}. In the case of qudits, we use one qudit for each lattice site, and map a spin state to a qudit state as
\begin{equation}
|\Ms\rangle = |\ell\rangle_{\quditd},
\end{equation}
where the subscript $\quditd$ ($=2\spin+1$) is used to label a qudit state, and $\ell=\Ms+\spin$ so that $\ell=0$ corresponds to $\Ms=-\spin$ and $\ell=2\spin$ corresponds to $\Ms=\spin$. 

\subsection{Qubit Hamiltonians and Suzuki-Trotter approximation}\label{subsec:Hamiltonians}
The qubit form of the spin Hamiltonian  \eqref{Eq:HeisenbergH} is expressed as 
\begin{equation}\label{Eq:Hq_general}
\Hamiltonian_{\rm q} = J\sum_{k=1}^{\NHq} h_k P_k,
\end{equation}
where $h_k$ is an expansion coefficient,  $P_k$, defined as
\begin{equation}
	P_k=\sigma_{\nqubittot-1}^k\sigma_{\nqubittot-2}^k\cdots \sigma_0^k, \qquad \sigma_n^k\in\{I,X,Y,Z\},
\end{equation}  is a direct product of Pauli matrices, and $\nqubittot=\Nsites\nqubit$ denotes the total number of qubits.
For the compact and direct mappings, $\Hamiltonian_{\rm q}$ can be derived by applying the methods in \cite{Sawayaetal2020} in a manner briefly described below. 
Taking the example of a two-site lattice for simplicity, we first write the spin Hamiltonian as
\begin{equation}
	\Hamiltonian = \sum_{\Ms'_1,\Ms'_0,\Ms_1,\Ms_0=-\spin}^{\spin}\eta_{\Ms'_1\Ms'_0\Ms_1\Ms_0}|\Ms'_1\Ms'_0\rangle\langle \Ms_1\Ms_0|,
\end{equation} 
where $\eta_{\Ms'_1\Ms'_0\Ms_1\Ms_0}$ are numerical values of the matrix elements derived by the application of Eqs.~\eqref{Eq:SzM} and \eqref{Eq:SpmM}.
Next, we convert $|\Ms'_1\Ms'_0\rangle\langle \Ms_1\Ms_0|$ to  a qubit operator form using the mappings
\eqref{Eq:CompactDef} or \eqref{Eq:DirectDef} and the identifications 
$|0\rangle\langle 0|=(I+Z)/2$, $|0\rangle\langle 1|=(X+iY)/2$, $|1\rangle\langle 0|=(X-iY)/2$, and 
$|1\rangle\langle 1|=(I-Z)/2$.
In the case of the direct mapping, we only need to consider the 
qubits whose state is changed by $|\Ms'_1\Ms'_0\rangle\langle \Ms_1\Ms_0|$. For example, for $\spin=1$, we map the matrix element $|\Ms'_1=-1,\Ms'_0=1\rangle\langle \Ms_1=0,\Ms_0=0|$ with $\Ms_1\ne \Ms'_1$ and $\Ms_0\ne \Ms'_0$ according to
\begin{align}
|{-1},1\rangle\langle 0,0|
={}&\ketq{001100}\braq{010010}
\nonumber\\
={}&\frac{1}{16}(X_4+iY_4)(X_3-iY_3)(X_2-iY_2)
\nonumber\\
&\quad\times(X_1+iY_1).
\end{align}

For the direct mapping, we find that the total number of terms $\NHq$ for a two-site lattice is 
proportional to $\spin^2$ and that 
$\NHq_{\rm direct}\approx 36\spin^2$ for $\spin$ in the range $1/2\le \spin\le 5$. For the compact mapping, we have not been able to analytically derive the scaling of $\NHq$, but as discussed in detail in Appendix \ref{Appendix_NtemrsHcompact}, we find empirically that $\NHqcompact\propto \spin^{2.4}$ holds approximately for
$\spin$ in the range 
$1/2\le \spin\le 63/2$.

In the case of the Dicke mapping, we simply replace each spin-$\spin$ operator $\boldSop{}$ in \eqref{Eq:HeisenbergH} by the total qubit spin operator 
$\boldSop{\text{q}}=\frac{\hbar}{2}\sum_{k=0}^{\nqubit-1}(X_k,Y_k,Z_k)$ 
\cite{Kawashima1994,Todo2001} so that the qubit Hamiltonian becomes
\begin{equation}\label{Eq:DickeH_qubit}
    \Hamiltonian_{\rm q}^{\rm Dicke} =
    \frac{J}{4}\sum_{(m,n)}
    \sum_{k=\kmin(m)}^{\kmax(m)}
    \sum_{l=\kmin(n)}^{\kmax(n)}
     \W_{kl},
\end{equation}
where
\begin{equation}\label{Eq:DefWkl}
	\W_{kl}=Z_kZ_l + Y_kY_l + X_kX_l.
\end{equation}
In Eq.~\eqref{Eq:DickeH_qubit}, $(m,n)$ labels a pair of interacting sites on the lattice and  $k$ and $l$ denote the indices of the qubits describing the spin at sites $m$ and $n$, respectively. The qubit index $k$ at site $n$ takes  values in the range 
$\kmin(n)\le k \le \kmax(n)$. For example, for a two-site $\spin=1$ system, the qubits 0 and 1 are assigned to site 0, meaning that $\kmin(0)=0$ and $\kmax(0)=1$, and qubits 2 and 3 are assigned to site 1, meaning that $\kmin(1)=2$ and $\kmax(1)=3$. The two-site, 
$\spin=1$ qubit Hamiltonian becomes
\begin{equation}\label{Eq:DickeH_qubitS1}
    \Hamiltonian_{{\rm q},\spin=1}^{\rm Dicke} =
    \frac{J}{4}
     \sum_{k=0}^1\sum_{l=2}^3
     \W_{kl}.
\end{equation}
The total number of terms $\NHq$ for a two-site lattice is 
$\NHq_{\text{Dicke}}=12\spin^2$.

In the case of the qudit mapping, we write the Hamiltonian in terms of generalized Gell-Mann matrices \cite{Luo2014},
\begin{equation}\label{Eq:QuditHamiltonian}
	\Hamiltonian_{d}=J\sum_{j=1}^Lg_k \Gamma_k,
\end{equation}
where the $g_k$'s are expansion coefficients and $\Gamma_k$, defined as
\begin{equation}
	\Gamma_k = \gamma^k_{\Nsites-1}\gamma^k_{\Nsites-2}\cdots \gamma^k_0, \qquad \gamma_n^k\in\{\GellMann_0,\ldots, \GellMann_{d^2-1}\},
\end{equation}  
is a direct product  of   $\quditd \times \quditd$ generalized Gell-Mann matrices $\GellMann_k$. The detailed definition  of $\GellMann_k$ is given in Appendix \ref{Appendix_GellMann}.
In laboratory experiments, implementation of the unitary operators $e^{i \theta \lambda_k}$  on a trapped-ion type quantum computer has been demonstrated in \cite{Ringbauer2022}. We find that for a two-site lattice, the number of terms in the Hamiltonian in the qudit mapping is the same as in the Dicke mapping, that is, 
$\NHq_{\text{qudit}}=\NHq_{\text{Dicke}}=12\spin^2$.

\begin{figure}
	\includegraphics[width=\figwidth]{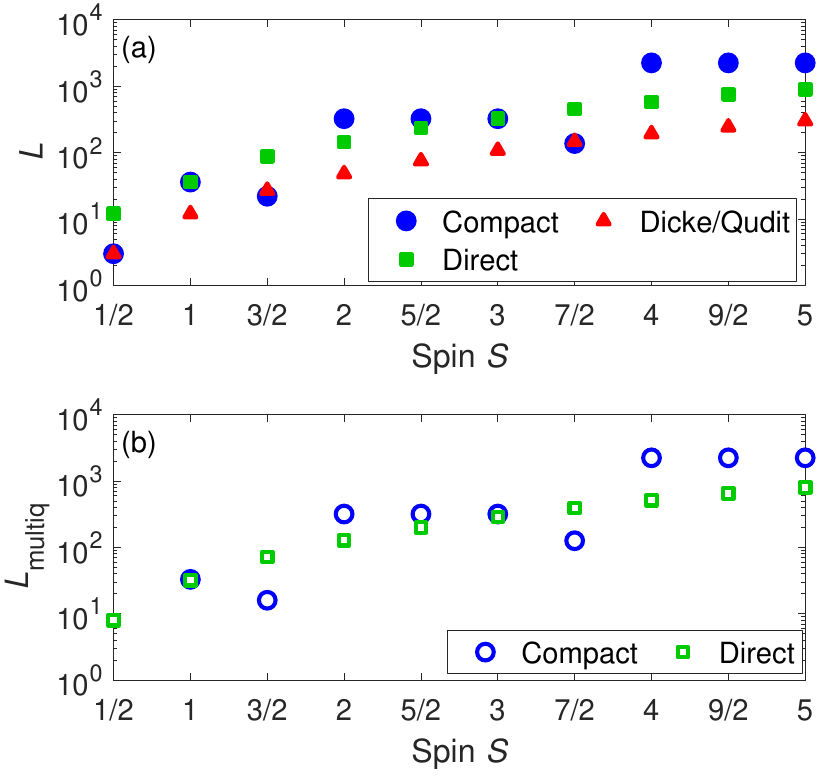}
	\caption{\label{Fig2}
		(a) Total number of terms $\NHq$ in the two-site Hamiltonian \eqref{Eq:Hq_general} for the compact, direct, Dicke, and qudit  mappings. Note that  
$\NHq_\text{Dicke}=\NHq_\text{qudit}$ and that   the vertical axis is in the logarithmic scale. (b) The number of multi-qubit terms in the Hamiltonian. The Dicke Hamiltonian \eqref{Eq:DickeH_qubit} and the qudit Hamiltonian \eqref{Eq:QuditHamiltonian} contain only two-qubit or two-qudit terms, and therefore, $\NHq^{\text{multiq}}_\text{Dicke}=\NHq^{\text{multiq}}_\text{qudit}=0$.
	}
\end{figure}

In Fig.~\ref{Fig2}(a), we show the total number of terms $\NHq$ in the two-site qubit Hamiltonian using the compact, direct, Dicke, and qudit mappings. We can see that the Dicke mapping almost always results in a smaller $\NHq$ than the compact or direct mappings. The only exception is when  $2\spin+1$ is exactly equal to  a power of 2, which occurs at $\spin=1/2$, $3/2$, and $7/2$ in Fig.~\ref{Fig2}. In these cases, all available qubit states correspond to spin states in the compact mapping, leading to a particularly compact representation of the qubit Hamiltonian \cite{Sawayaetal2020}. However, even if the total number of terms $\NHq$ is approximately the same in the Dicke and compact mappings for $\spin=1/2$, $3/2$, and $7/2$, the Dicke mapping Hamiltonian \eqref{Eq:DickeH_qubit} contains only two-qubit operators, while the compact mapping Hamiltonian contains many  operators acting on more than two qubits. 
For example, at $\spin=3/2$, the compact-mapping qubit Hamiltonian contains $\NHqcompact=22$ terms:  six  two-qubit operators, eight three-qubit operators, and eight  four-qubit operators. At $\spin=2$, we have $\NHqcompact=324$ terms, out of which six are two-qubit operators, 28 are three-qubit operators, 73 are four-qubit operators, 118 are five-qubit operators, and 99 are six-qubit operators. In Fig.~\ref{Fig2}(b), we show the 
number $\NHq_{\text{multiq}}$ of multi-qubit terms in the Hamiltonian, where a multi-qubit term is defined as an operator which acts on more than two qubits (or qudits in the case of the qudit mapping).
The reason for the absence of many-qubit operators acting on three or more qubits in   the two-site qubit Dicke Hamiltonian is  that the single-site spin operators such as $\Sop{z}{n}$ and $\Sop{\pm}{n}$ are expressed as permutation-symmetric sums of single-qubit operators. This representation of $\Sop{z}{n}$ and $\Sop{\pm}{n}$   is only possible when the single-site  large-$\spin$ states are represented as permutation-symmetric qubit states.

In order to propagate the wave function forward in time, we adopt the simplest form of the Suzuki-Trotter approximation \cite{Trotter1959,Suzuki1976},
\begin{equation}\label{Eq:SuzukiTrotterApprox}
    |\psi(t=\Ntrotter \dt)\rangle \approx  \left[ \UST(\dtau) \right]^{\Ntrotter}|\psi_0\rangle,
\end{equation}
where $|\psi_0\rangle$ is the initial state, $\Ntrotter$ is the number of Suzuki-Trotter steps,
\begin{equation}\label{Eq:UST_qubit}
    \UST(\dtau)=e^{-i \dtau h_\NHq P_\NHq} 
    e^{-i \dtau h_{\NHq-1} P_{\NHq-1}}\cdots 
    e^{-i \dtau h_1 P_1}
\end{equation}
is the unitary operator of one Suzuki-Trotter step,
and $\dtau = J\dt/\hbar$ is the dimensionless time step. For the qudit mapping, an expression equivalent to Eq.~\eqref{Eq:UST_qubit} is used where the qubit operators $P_k$ are replaced by qudit operators $\Gamma_k$. 

\begin{figure*}
	\includegraphics[width=1\textwidth]{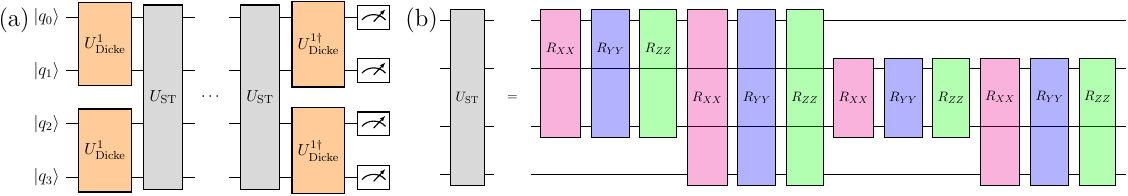}
	\caption{\label{Fig3}
		(a) Structure of the Suzuki-Trotter circuit for the Dicke mapping of a two-site, $\spin=1$ system. The qubits are initialized in the state $|\psi_0\rangle=|q_3 q_2 q_1 q_0\rangle$.
		(b) Structure of one Suzuki-Trotter step $\UST$. The two-qubit rotation gates $R_{\alpha\alpha}$ with $\alpha=X$, $Y$, or $Z$ are defined as 
		$R_{\alpha\alpha}=\exp\left(-i\frac{\dtau}{4}\alpha\alpha \right)$. Each  gate $R_{\alpha\alpha}$ operates on the two qubits covered by the colored rectangle.
	}
\end{figure*}

In the case of the Dicke mapping, it is necessary to add  the Dicke operator $\UDicke^{\spin}$ before the Suzuki-Trotter time evolution and $\UDicke^{\spin\dagger}$ before the final measurement in order to transform a computational basis state (a state $\ketq{q_{\nqubittot-1}\cdots q_0}$ with 
$q_k\in \{0,1\}$) to a Dicke state and back to the computational basis. We illustrate the structure of the Suzuki-Trotter circuit for the Dicke mapping in Fig.~\ref{Fig3}. It can be seen in Fig.~\ref{Fig3} that, although the Dicke mapping is compact, it requires the coupling of non-neighboring qubits. This means that quantum computers having an all-to-all qubit connectivity, such as trapped-ion  \cite{Wright2019,Pino2021,Moses2023_PRX} and  neutral atom  \cite{Bluvstein2023} devices are particularly well suited for  the practical realization of the Dicke mapping.
On the other hand, quantum computers constructed from qubits realized as superconducting  circuits are restricted to nearest-neighbor qubit connectivity, which  means that superconducting quantum computers are not suited for the realization of the Dicke mapping because of the large increase of the number of two-qubit gates required to implement the 
coupling of non-neighboring qubits.

\section{Results}\label{Sec:Results}

\subsection{Comparison of compact, direct, Dicke, and qudit mappings}\label{subsec:ComparisonMappings}
In order to assess the performance of the different mappings introduced in Secs.~\ref{subsec:QubitMappings} and \ref{subsec:Hamiltonians} in the presence of noise, we simulate the spin dynamics of a two-site lattice described by the Heisenberg Hamiltonian \eqref{Eq:HeisenbergH}. We conduct the simulation at two values of the spin $\spin$, i.e., $\spin=1$ and $\spin=3/2$. The time-dependent wave function $|\psi(t)\rangle$ is calculated according to Eq.~\eqref{Eq:SuzukiTrotterApprox} with a fixed time step $\dtau=0.2$ and the number of Suzuki-Trotter steps varying in the range $0\le \Ntrotter \le 31$, corresponding to a time range of $0\le Jt/\hbar \le 6.2$. 
The time step size $\dtau=0.2$ is selected so that $\dtau$ is significantly smaller than 1, which is required in the Suzuki-Trotter approximation \eqref{Eq:SuzukiTrotterApprox}. At the same time, the number of time steps ($\le 31$) is small enough so that the error due to the noise is at an acceptable level when the simulation is conducted using the H1-1 device.
The same $\dtau$ is selected for both $\spin=1$ and $\spin = 3/2$ to allow a consistent comparison.

We adopt the initial wave function $|\psi(t=0)\rangle = |\Ms_1=-\spin,\Ms_0=\spin\rangle$ and evaluate the time-dependent population $p_0(t)$  in the initial state,
\begin{equation}
	p_0(t)=|\langle \psi(0)|\psi(t) \rangle |^2.
\end{equation}
The initial-state population $p_0(t)$ depends on time because the initial state $|{-\spin},\spin\rangle$ is not an eigenstate of 
$\Hamiltonian$.

For the compact, direct, and Dicke mappings, we carry out the simulation using the Quantinuum H1-1 trapped-ion type quantum computer \cite{QuantinuumH11_2024}, which has 20 qubits realized by 20 trapped ${}^{171}$Yb$^+$ ions. The qubit states $|0\rangle$ and $|1\rangle$ are implemented as the states  
${}^2S_{1/2}(F=0,m_F=0)$ and ${}^2S_{1/2}(F=1,m_F=0)$ of Yb$^+$  \cite{Pino2021}. At the time of the simulations (July, August and October, 2024), the single and two-qubit gate errors  $\erroroneq$ and $\errortwoq$ of H1-1 were $\erroroneq\approx2\times 10^{-5}$ and $\errortwoq\approx1\times 10^{-3}$, respectively \cite{QuantinuumH11_2024}. We also use the H1-1 emulator, which implements an error model \cite{RyanAnderson2021} closely 
mimicking the noise in the real H1-1 device. The parameters of the H1-1 emulator can be found in the documentation available at \cite{QuantinuumH11E_2024}. In the simulations using the H1-1  (real device and emulator), we  repeat the execution of each circuit $\Nshots=1024$ times in order to estimate the populations.

For the qudit mapping, we  employ a noise model in which the qudit density matrix $\rho$ after each two-qudit gate is subject to a completely depolarizing noise channel
\begin{equation}
	\rho \to \rho' = (1-\errortwoq)\rho + \frac{\errortwoq}{\quditd^2}I_{\quditd^2\times\quditd^2},
\end{equation}
where $I_{\quditd^2\times\quditd^2}$ is the identity matrix. We conduct simulations using two values of the two-qudit gate error, $\errortwoq=1\times 10^{-3}$, which equals  the two-qubit gate error of the Quantinuum H1-1  \cite{QuantinuumH11_2024}, and $\errortwoq=1\times 10^{-2}$, which is a more realistic assumption of the two-qudit gate error achievable in the present qudit systems. Note that a gate error of 
$\errortwoq=3\times 10^{-2}$ was reported in \cite{Ringbauer2022} for the two-qudit M{\o}lmer-S{\o}rensen gate.
 
\begin{figure}
	\includegraphics[width=\figwidth]{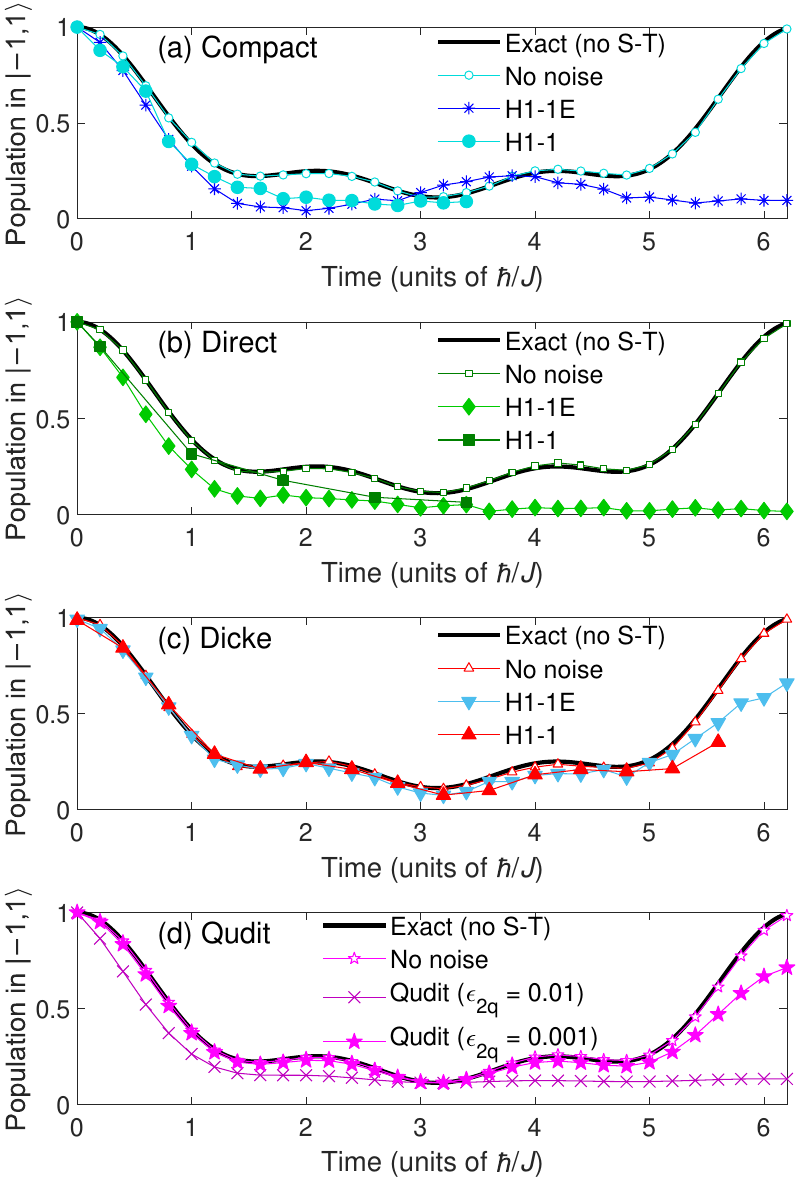}
	\caption{\label{Fig4}
		Populations in the $|\Ms_1,\Ms_0\rangle=|{-1},1\rangle$ state for the Suzuki-Trotter simulation of  a
		two-site, $\spin=1$ lattice obtained using the (a) compact, (b) direct, (c) Dicke, and (d) qudit  mappings. The initial state is $|\psi(0)\rangle =|{-1},1\rangle$. In all plots, the exact populations obtained without using the Suzuki-Trotter approximation are drawn by black solid lines and the populations obtained by the Suzuki-Trotter approximation in the absence of  noise are shown with open symbols. The curves in (a)--(c) labeled by ``H1-1'' are obtained by the Quantinuum H1-1 trapped-ion quantum computer and the curves labeled by ``H1-1E'' are obtained by using the H1-1 emulator.
		The standard deviation  estimated by bootstrapping \cite{EfronTibshirani1994} is smaller than 0.02 for all data points in all panels. Error bars are not shown because they are smaller than the curve symbols.
	}
\end{figure}

In Fig.~\ref{Fig4}, we show the initial-state population $p_0(t)$ for an $\spin=1$ two-site lattice, obtained by the compact, direct, Dicke, and qudit mappings. For reference, the explicit expressions of the qubit and qudit Hamiltonians are given in Appendix \ref{Appendix_quditHamiltoniansS1}. The number of qubits employed for the two sites is $2\nqubit=4$ in the compact and Dicke mappings, and $2\nqubit=6$ in the direct mapping. When the circuits are transpiled (using {\sc tket} \cite{Sivarajah2020}), the total number of $R_{ZZ}$ gates (H1-1's native two-qubit gate) for one Suzuki-Trotter step is $\Ntwoq=63$ in the direct mapping, $\Ntwoq=56$ in the compact mapping, and 
$\Ntwoq=16$ (including $\UDicke^{1}$ and $\UDicke^{1\dagger}$) in the Dicke mapping.  
In the qudit mapping, the number of two-qutrit gates is the same as the number of terms in the Hamiltonian,
$\Ntwoq=\NHq_{\rm Dicke}=12$.
 
We can see in Figs.~\ref{Fig4}(a) and \ref{Fig4}(b) that in the case of the compact and direct mappings, $p_0(t)$ obtained using H1-1 starts deviating from $p_0(t)$  obtained in the absence of noise already in the early propagation time range of  $Jt/\hbar>0.2$. On the other hand, $p_0(t)$ obtained using the Dicke and qudit mapping (with the smaller two-qutrit gate error $\errortwoq=1\times 10^{-3}$) shown in Fig.~\ref{Fig4}(c) and \ref{Fig4}(d) agree well with the ``No noise'' curves. If we define the average error in the population as 
\begin{equation}\label{Eq:AverageErrorDef}
	\overline{\varepsilon}=\frac{1}{N_{\rm p}}\sum_{t_n\le T}|p_0^{\text{(noisy)}}(t_n)
	-p_0^{\text{(no noise)}}(t_n)|,
\end{equation}
where $t_n$ ($n=1,\ldots,N_{\rm p}$) are the time instants for the data shown in Fig.~\ref{Fig4}, we obtain
$\overline{\varepsilon}_{\text{Dicke}}(\text{H1-1})=0.012$ and 
$\overline{\varepsilon}_{\text{qudit}}(\errortwoq=0.001)=0.011$ for $T=3.2\hbar/J$, much smaller than 
$\overline{\varepsilon}_{\text{compact}}(\text{H1-1})=0.080$ and 
$\overline{\varepsilon}_{\text{direct}}(\text{H1-1})=0.060$.
The average error $\overline{\varepsilon}_{\text{qudit}}(\errortwoq=0.01)=0.081$
obtained in the qudit simulation using the larger two-qutrit gate error of $\errortwoq=0.01$ is comparable to the average errors obtained by   the compact and direct mappings.
We note that the agreement between the populations obtained with no noise and the populations obtained using the Dicke and qudit mappings [Figs.~\ref{Fig4}(c) and \ref{Fig4}(d)] could be improved by error mitigation such as postselection. On the other hand,   the populations obtained using the compact and direct mappings  cannot be meaningfully improved by error mitigation because the error due to the noise is too large.
We also learn from Fig.~\ref{Fig4} that the H1-1E emulator provides an accurate error model capable of reproducing the results obtained on the real H1-1 device.

\begin{figure}
	\includegraphics[width=\figwidth]{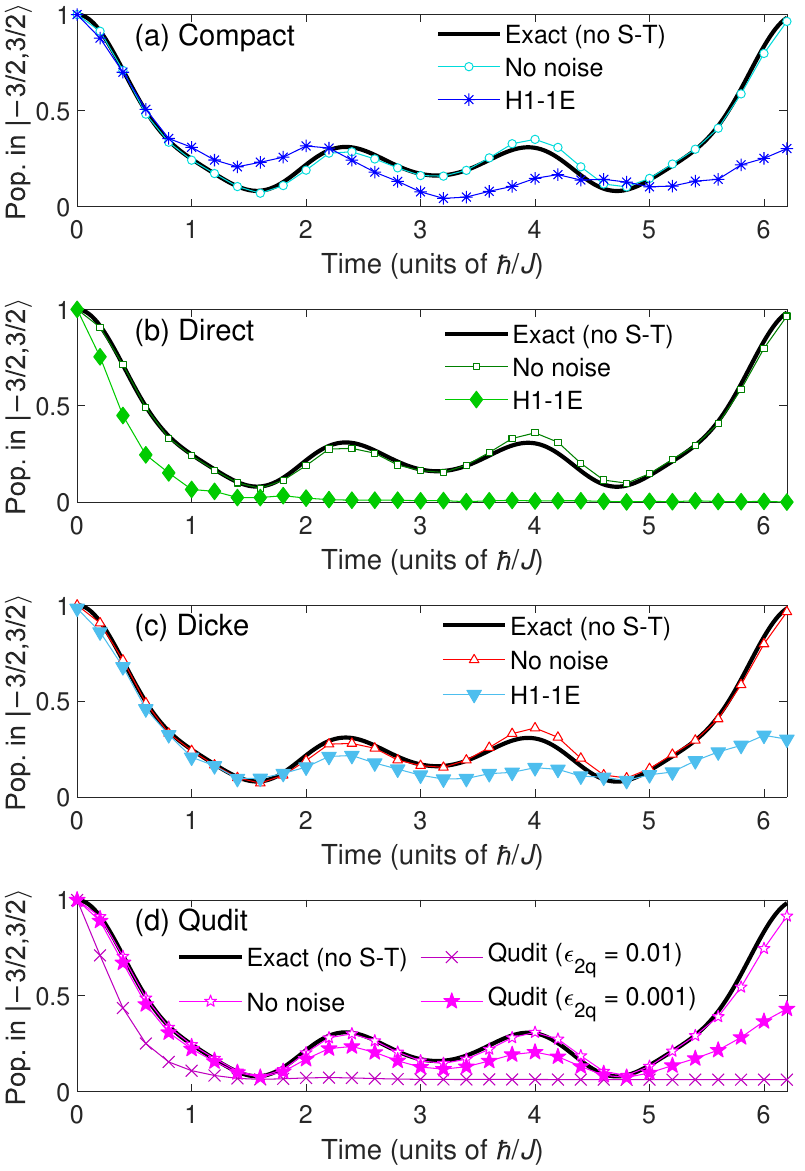}
	\caption{\label{Fig5}
		Populations in the $|\Ms_1,\Ms_0\rangle=|{-3/2},3/2\rangle$ state for the Suzuki-Trotter simulation of  a
		two-site, $\spin=3/2$ lattice obtained using the (a) compact, (b) direct, (c) Dicke, and (d) qudit mappings.
		The initial state is $|\psi(0)\rangle =|{-3/2},3/2\rangle$. 
		The exact populations obtained without using the Suzuki-Trotter approximation are plotted by black solid lines and the populations  obtained by the Suzuki-Trotter approximation in the absence of noise are plotted with open symbols. The curves in (a)--(c) labeled by  ``H1-1E'' are obtained using the Quantinuum H1-1 emulator. 
		The standard deviation estimated by bootstrapping \cite{EfronTibshirani1994} is smaller than 0.02 for
all data points in all panels. Error bars are not shown because they are smaller than the curve symbols.}
\end{figure}

In order to compare the different mappings when $\NHqcompact<\NHq_{\rm Dicke}$, we show $p_0(t)$ in Fig.~\ref{Fig5} for a two-site lattice with $\spin=3/2$
in which the number of terms in the Hamiltonians is $\NHqcompact=22$ and $\NHq_{\rm Dicke}=27$. The qubit Hamiltonians are shown in Appendix 
\ref{Appendix_quditHamiltoniansS3half}. The number of qubits is  $2\nqubit=4$ in the compact mapping, $2\nqubit=6$ in the Dicke mapping, and $2\nqubit=8$ in the direct mapping. After transpiling the Suzuki-Trotter circuits to 
the native gates of H1-1 ($R_{ZZ}$, $R_X$, and $R_Z$), the number of two-qubit gates for one Suzuki-Trotter step  becomes $\Ntwoq=41$ in the compact mapping, $\Ntwoq=143$ in the direct mapping, and $\Ntwoq=49$ in the Dicke mapping. In the qudit mapping, we have 
$\Ntwoq=\NHq_{\rm Dicke}=27$ two-ququart gates. We use the emulator H1-1E to obtain the results shown in Fig.~\ref{Fig5}. 
Considering the comparisons shown in Fig.~\ref{Fig4}, we expect that results obtained here using H1-1E are close to the results to be obtained using the real H1-1 device.

We find in Fig.~\ref{Fig5} that, similarly to the case of $\spin=1$, $p_0(t)$ obtained using the  Dicke and qudit ($\errortwoq=0.001$) mappings  shown in Figs.~\ref{Fig5}(c) and \ref{Fig5}(d) deviate only slightly from the population obtained with noise-free conditions. 
By setting $T=3.2\hbar/J$ in Eq.~\eqref{Eq:AverageErrorDef}, we obtain 
$\overline{\varepsilon}_{\text{Dicke}}(\text{H1-1E})=0.036$ and 
$\overline{\varepsilon}_{\text{qudit}}(\errortwoq=0.001)=0.030$ for the average errors.
The population obtained using the direct mapping shown in Fig.~\ref{Fig5}(b) and the qudit mapping with  $\errortwoq=0.01$ shown in Fig.~\ref{Fig5}(d) rapidly decrease respectively to the values $p_0(Jt/\hbar>2)\approx 1/2^{8}\approx 0.039$ for the direct mapping and $p_0(Jt/\hbar>2)\approx 1/4^{2}\approx 0.063$ for the qudit mapping with $\errortwoq=0.01$. This implies 
that a completely mixed state, in which all states are equally populated, is produced because of the error accumulation, so that
the value of $p_0$ is given by the inverse of the total number of qubit or qudit states. Consequently, the large average errors of 
$\overline{\varepsilon}_{\text{direct}}(\text{H1-1E})=0.16$ and
$\overline{\varepsilon}_{\text{qudit}}(\errortwoq=0.01)=0.14$ 
are obtained.
Because of the small number of terms in the Hamiltonian, the simulation employing the compact mapping results in rather accurate values of $p_0(t)$ as can be seen in Fig.~\ref{Fig5}(a). We obtain 
$\overline{\varepsilon}_{\text{compact}}(\text{H1-1E})=0.069$, which is only about a factor of two larger 
than $\overline{\varepsilon}_{\text{Dicke}}(\text{H1-1E})$. 

From the comparison of the results obtained by the different mappings shown in Figs.~\ref{Fig4} and \ref{Fig5}, we conclude that 
both the Dicke mapping and the qudit mapping give equally accurate results, provided that a small two-qudit gate error of $\errortwoq=0.001$ can be achieved. This suggests that there is no particular  advantage of simulating large-$\spin$ systems using qudit devices. On the other hand, the Dicke mapping is an efficient way of simulating large-$\spin$ lattices using quantum computers having an all-to-all qubit connectivity such as trapped-ion   \cite{Moses2023_PRX} and neutral atom  \cite{Bluvstein2023} devices.

\subsection{Dicke mapping for \boldmath{$1/2\le\spin\le 5/2$}}\label{subsec:DickeMappingLargeS}
In order to assess the performance of the Dicke mapping for a larger number of qubits, we simulate  a four-site open-ended linear lattice, meaning that the sum over $(m,n)$ in Eq.~\eqref{Eq:HeisenbergH} is taken over $(m,n)=(0,1)$, $(1,2)$ and $(2,3)$  for $\spin=1/2$, $1$, $3/2$, $2$, and $5/2$. The simulations are run on the Quantinuum H1-1 trapped-ion quantum computer \cite{QuantinuumH11_2024} using $\Nshots=1024$. Because $\nqubit=2\spin$ qubits per site are required in the Dicke mapping, all the 20 qubits of H1-1 are employed in the $\spin=5/2$ simulation. 
The main purpose of the simulations presented in this section  is to consistently compare the performance of the Dicke mapping for different values of $\spin$ and to investigate the extent of the error due to the noise as the number  of Suzuki-Trotter time steps increases. We set  the time step adopted in the Suzuki-Trotter approximation \eqref{Eq:UST_qubit} 
to be $\dtau=\pi/10\approx 0.314$. The number of employed Suzuki-Trotter time steps  is proportional to $t$ according to $\Ntrotter=Jt/(\hbar \dtau)$.

After transpilation with {\sc tket} \cite{Sivarajah2020}, 
the number of two-qubit gates in the quantum circuits for one Suzuki-Trotter step (including $\UDicke^{\spin}$ and $\UDicke^{\spin\dagger}$) becomes
$\Ntwoq(\spin=1/2)=8$, $\Ntwoq(\spin=1)=44$, $\Ntwoq(\spin=3/2)=125$, $\Ntwoq(\spin=2)=252$, and 
$\Ntwoq(\spin=5/2)=425$ for the different values of $\spin$.
The longest circuit is a 20-qubit circuit for $\spin=5/2$ with  $\Ntrotter=11$ Suzuki-Trotter steps ($Jt/\hbar = 3.46$), resulting in $\Ntwoq=2675$ two-qubit gates.

\begin{figure}
	\includegraphics[width=0.42\paperwidth]{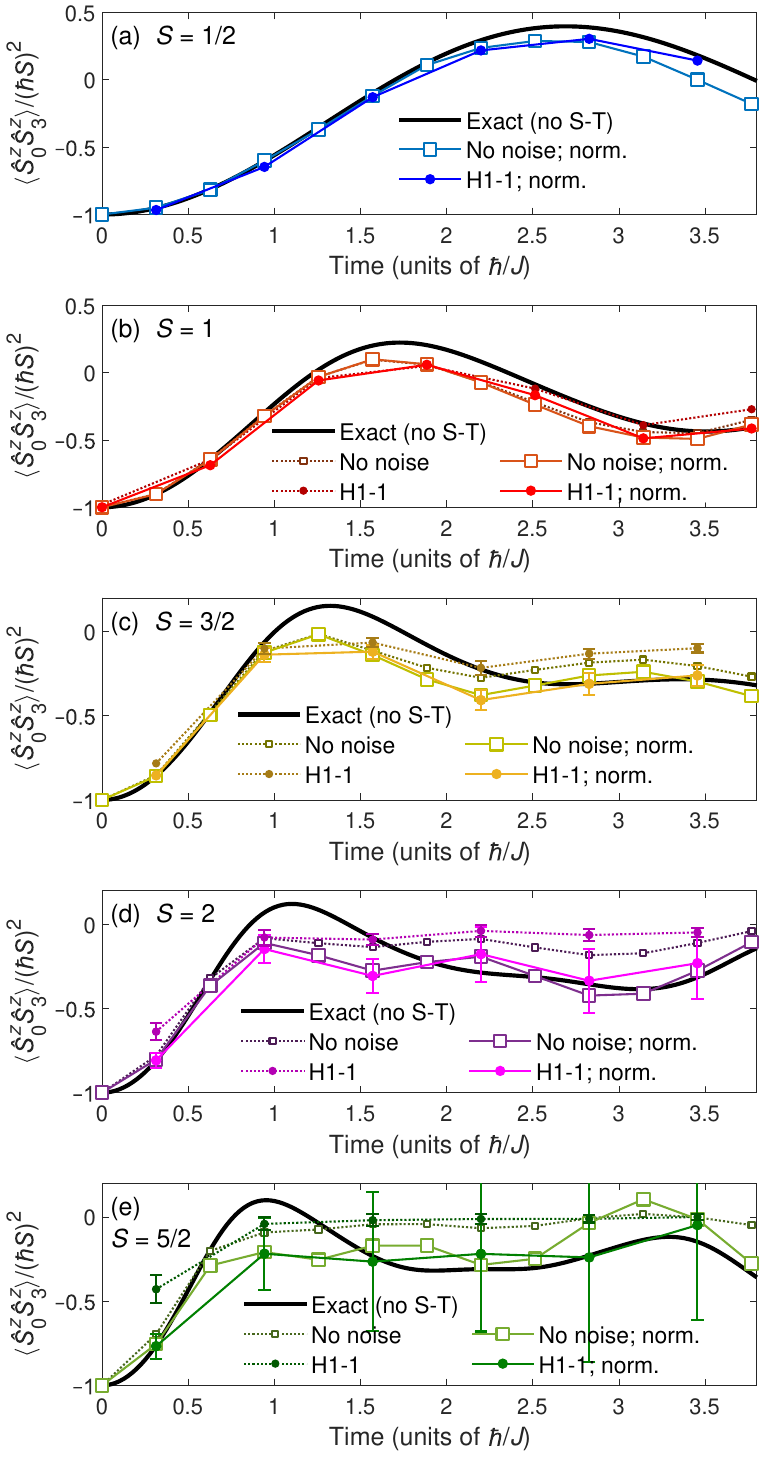}
	\caption{\label{Fig6} 
		The time-dependent expectation value  $\langle \Sop{z}{0}\Sop{z}{3}\rangle (t)$ for a  four-site linear lattice defined by  Eq.~\eqref{Eq:Sz0Sz3Def}, for (a) $\spin=1/2$, 
		(b) $\spin=1$, (c) $\spin=3/2$, (d) $\spin=2$, and (e) $\spin=5/2$.
		We show the exact results (labeled ``Exact (no S-T)''), the results obtained using the Suzuki-Trotter approximation in the absence of  noise (``No noise''), and the results obtained using the Quantinuum H1-1. Curves labeled by ``norm'' are obtained after the postselection and normalization defined by Eq.~\eqref{Eq:S0SN_renorm}. Because  the effect of the postselection and normalization 
is small in (a) for both the results obtained without noise and the results obtained using H1-1, (|$\langle \Sop{z}{0}\Sop{z}{3}\rangle - \langle \Sop{z}{0}\Sop{z}{3}\rangle_{\rm norm}|<0.03$), we show only the curves after the postselection and normalization. Note also that 
the ``No noise'' and ``No noise; norm.'' curves in (b) almost completely overlap.   The error bars represent one standard deviation and are shown only when the standard deviation exceeds $0.03$. 
	}
\end{figure}

In Fig.~\ref{Fig6}, we plot the time-dependent two-site observable 
\begin{equation}\label{Eq:Sz0Sz3Def}
	\langle \Sop{z}{0}\Sop{z}{3}\rangle (t) =\langle \psi(t)|\Sop{z}{0}\Sop{z}{3}|\psi(t)\rangle,
\end{equation}
where $|\psi(t)\rangle$ is the time-dependent wave function evaluated using the Suzuki-Trotter approximation \eqref{Eq:SuzukiTrotterApprox} with $\dtau=\pi/10\approx 0.314$. The  initial state is
\begin{equation}\label{Eq:initialstate}
	|\psi(t=0)\rangle = |\Msi{3}=-\spin,\Msi{2}=-\spin,\Msi{1}=-\spin,\Msi{0}=\spin\rangle.
\end{equation} 
In order to facilitate the comparison 
of $\langle \Sop{z}{0}\Sop{z}{3}\rangle$ evaluated for the different values of $\spin$, we show the scaled quantity $\langle \Sop{z}{0}\Sop{z}{3}\rangle /(\hbar\spin)^2$ in Fig.~\ref{Fig6}.

In practice, we evaluate 
$\langle \Sop{z}{0}\Sop{z}{3}\rangle (t)$ according to 
\begin{equation}\label{Eq:S0SN_practicedef}
	\langle \Sop{z}{0}\Sop{z}{3}\rangle \approx \sum_{\vec{\Ms}} \Ms_0\Ms_3 \frac{N_{\vec{\Ms}}}{\Nshots},
\end{equation}
where  $N_{\vec{\Ms}}$ is the number of measurements of a qubit state representing a spin state $|\vec{\Ms}\rangle$. This means that the sum in Eq.~\eqref{Eq:S0SN_practicedef} is taken only over qubit states which are mapped to the spin states by $\UDicke^{\spin}$. For example, for $\spin=1$, the sum in
Eq.~\eqref{Eq:S0SN_practicedef} includes the eight-qubit states $|Q_3Q_2Q_1Q_0\rangle$ with 
$Q_i\in \{00, 01, 11\}$.


%

We also show in Fig.~\ref{Fig6} the result of a simple form of error mitigation by the postselection and  normalization. 
The postselection is a widely used scheme for error mitigation  \cite{BonetMonroig2018,McArdleYuanBenjamin2019, Cai2023}, 
in which results of the measurements are discarded when  the symmetry of the Hamiltonian is violated  or  the measured bit string is not included  in the qubit encoding. Because of the error in the Suzuki-Trotter approximation \eqref{Eq:SuzukiTrotterApprox}, 
the total population in the qubit states representing Dicke states decreases and becomes smaller than one even when noise is absent. The only exception is  $\spin=1/2$, in which the Dicke state population is conserved because $\nqubit=1$, that is, all qubit states represent a Dicke state.  In addition,
because of the noise, the total magnetic quantum number 
\begin{equation}
\Ms_{\rm tot}=\sum_{n=0}^{\Nsites -1} M_n
\end{equation}
is not conserved during the simulation, even though the  Hamiltonian 	
$\Hamiltonian$ commutes with $\Sop{z}{\text{tot}}=\sum_{n=0}^{\Nsites -1}\Sop{z}{n}$. Therefore, even if we start the simulation from the intial state \eqref{Eq:initialstate} having $\Ms_{\rm tot}=-2\spin$, the total population summed over all states having the same $\Ms_{\rm tot}$ decreases during the Suzuki-Trotter simulation. A simple way of error mitigation is to post-select only the spin states having the same $\Ms_{\rm tot}$ as the initial state and  normalize their population to one before calculating the expectation value,
\begin{equation}\label{Eq:S0SN_renorm}
	\langle \Sop{z}{0}\Sop{z}{3}\rangle_{\text{norm}} = \sum_{\vec{\Ms}:\Ms_{\rm tot}=\Ms^{\text{i}}_{\rm tot}} \Ms_0\Ms_3 \frac{N_{\vec{\Ms}}}{N_{\Ms^{\text{i}}_{\rm tot}}},
\end{equation}
where $\Ms^{\text{i}}_{\rm tot}$ is the total magnetic number of the initial state ($=2\spin$ in Fig.~\ref{Fig6}), and $N_{\Ms^{\text{i}}_{\rm tot}}$ is the total number of measurements of qubit states corresponding to the  spin states whose total magnetic quantum numbers equal  
$\Ms^{\text{i}}_{\rm tot}$.
The error-mitigated values of $\langle \Sop{z}{0}\Sop{z}{3}\rangle$ evaluated according to Eq.~\eqref{Eq:S0SN_renorm} are shown with solid lines (labeled ``norm'') in Fig.~\ref{Fig6}.
An alternative scheme to mitigate  the decrease in the population in the Dicke states using projection operators was suggested in \cite{Maskara2025}. The conservation
 of the population can also be improved by adopting more complex Suzuki-Trotter schemes similar to those reported in \cite{Childs2021,Morales2025}.

For reference, we give the total population  $p_{\Ms_{\rm tot}}$ in the states with 
$\Ms_{\rm tot}=\Ms^{\text{i}}_{\rm tot}$  estimated by
$p_{\Ms_{\rm tot}}=N_{\Ms^{\text{i}}_{\rm tot}}/\Nshots$  at the largest $t=t_{\rm max}$ in the H1-1 simulations ($Jt_{\rm max}/\hbar = 3.77$ for $\spin=1$, and $Jt_{\rm max}/\hbar=3.46$ for $\spin\ne 1$). 
In the noise-free simulations, we obtain 
$p_{\Ms_{\rm tot}}^{\spin=1/2}(t_{\rm max})=1$,
$p_{\Ms_{\rm tot}}^{\spin=1}(t_{\rm max})=0.909$,
$p_{\Ms_{\rm tot}}^{\spin=3/2}(t_{\rm max})=0.691$,
$p_{\Ms_{\rm tot}}^{\spin=2}(t_{\rm max})=0.402$,
and
$p_{\Ms_{\rm tot}}^{\spin=5/2}(t_{\rm max})=0.161$, and, in
 the simulations conducted using H1-1, we obtain
$p_{\Ms_{\rm tot}}^{\spin=1/2}(t_{\rm max})=0.959$,
$p_{\Ms_{\rm tot}}^{\spin=1}(t_{\rm max})=0.665$,
$p_{\Ms_{\rm tot}}^{\spin=3/2}(t_{\rm max})=0.331$,
$p_{\Ms_{\rm tot}}^{\spin=2}(t_{\rm max})=0.0898$,
and
$p_{\Ms_{\rm tot}}^{\spin=5/2}(t_{\rm max})=0.0195$. The number of measurements discarded after the postselection is equal to $\Nshots(1-p_{\Ms_{\rm tot}}^{\spin}(t_{\rm max}))$, where $\Nshots=1024$.
Because of the small values of 
$p_{\Ms_{\rm tot}}^{\spin}(t_{\rm max})$ for large $\spin$, the error bars of the error-mitigated H1-1 curves become large, as can be seen for  $\spin=5/2$  in Fig.~\ref{Fig6}(e).

We can see in Fig.~\ref{Fig6} that $\langle \Sop{z}{0}\Sop{z}{3}\rangle$ obtained using H1-1 agrees reasonably well with the noise-free curves for $\spin<5/2$. 
We also find that the results obtained using H1-1 become closer to the noise-free results after the error mitigation by the postselection and normalization.
 For example, at $\spin=2$, we have the average discrepancy $\overline{d}(\spin=2)= 0.070$, where $\overline{d}$ is defined as 
\begin{equation}\label{Eq:Defaveragediscrepancy}
\overline{d}=\frac{1}{N_{\rm p} \hbar^2\spin^2}\sum_{n}\left|\langle \Sop{z}{0}\Sop{z}{3}\rangle_{\text{no noise}}(t_n)-\langle \Sop{z}{0}\Sop{z}{3}\rangle_{\text{H1-1}}(t_n) \right|,
\end{equation}
where the sum is taken over the time points $t_n$ at which the H1-1 data are obtained and $N_{\rm p}$ is the number of data points. After the error mitigation by normalization of  both the noise-free and the H1-1 data, $\overline{d}$ decreases to $\overline{d}_{\text{norm.}}(\spin=2)= 0.037$, showing that the error mitigation improves 
the agreement of the results obtained using H1-1 with the noise-free results.
For the largest spin considered, $\spin=5/2$, accurate values of $\langle \Sop{z}{0}\Sop{z}{3}\rangle (t)$ are obtained only 
at the Suzuki-Trotter steps of  $\Ntrotter =1$ ($Jt/\hbar=0.314$) and $3$ ($Jt/\hbar=0.942$).
At the larger $\Ntrotter$ values, the effect of the noise is so large that meaningful results cannot be obtained even after the  error mitigation. 
Because of the relatively large time step $\dtau =0.314$,  the Suzuki-Trotter approximation is less accurate for $\spin\ge 3/2$, 
leading to a substantial deviation of the "No noise" curve from the "Exact" curve as can be seen in  Fig.~\ref{Fig6}.

It can be seen in Fig.~\ref{Fig6} that the initial increase of  $\langle \Sop{z}{0}\Sop{z}{3}\rangle /(\hbar\spin)^2$ 
in time is more rapid when $\spin$ takes a larger value. This observation can be explained by second-order perturbation theory. After some algebra, we 
derive for small $Jt/\hbar\ll 1$
\begin{equation}\label{Eq:PT2_S0Sn}
	\frac{\langle \Sop{z}{0}\Sop{z}{3}\rangle(t)}{\hbar^2\spin^2}
    \approx \frac{\langle \psi_{\rm PT2}(t)| \Sop{z}{0}\Sop{z}{3}|\psi_{\rm PT2}(t)\rangle}{\hbar^2\spin^2}
    =-1+\spin\left(\frac{Jt}{\hbar}\right)^2,	
\end{equation}
where the wave function up to the second order in $t$ is calculated according to 
\begin{equation}
	|\psi_{\rm PT2}(t)\rangle = \left(1-\frac{i t}{\hbar}\Hamiltonian -\frac{t^2}{2\hbar^2}\Hamiltonian^2\right)|-\spin,-\spin,-\spin,\spin\rangle.
\end{equation}
Equation \eqref{Eq:PT2_S0Sn} shows that the second time derivative (the acceleration) of 
$\langle \Sop{z}{0}\Sop{z}{3}\rangle/(\hbar\spin)^2$ is proportional to $\spin$ for small $t$, which explains the faster rise  of $\langle \Sop{z}{0}\Sop{z}{3}\rangle/(\hbar\spin)^2$ in time for the larger $\spin$.

\subsection{Time step size in the Suzuki-Trotter approximation}\label{subsec:TimeStepSuzukiTrotter}
As  shown in Fig.~\ref{Fig6}, the deviation of the noise-free results obtained using the Suzuki-Trotter approximation from the exact results obtained without using the Suzuki-Trotter approximation  increases as $\spin$ increases, suggesting that a smaller value of $\dtau$ than 0.314 needs to be selected for large $\spin$.

In order to investigate  how small $\dtau$ should be for larger values of $\spin$, we evaluate the final spin-state population of a two-site lattice 
at a final time $Jt/\hbar =1$ with the different values of Suzuki-Trotter steps $\Ntrotter$,
\begin{equation}\label{Eq:pMfMiteqone}
p_{\vec{\Ms}^{\rm f}\vec{\Ms}^{\rm i}}(t=\hbar/J)=
\left|\langle\vec{\Ms}^{\rm f} | 
\left[ \UST(\dtau) \right]^{\Ntrotter}
|\vec{\Ms}^{\rm i} \rangle \right|^2,
\end{equation}
where $|\vec{\Ms}^{\rm i}\rangle=|\Ms^{\text{i}}_{1},\Ms^{\text{i}}_{0}\rangle$ and 
$|\vec{\Ms}^{\text{f}}\rangle=|\Ms^{\text{f}}_{1},\Ms^{\text{f}}_{0}\rangle$ are the initial and final spin states, respectively, 
and the time step $\dtau$ is defined as the inverse of the number of Suzuki-Trotter time steps,
\begin{equation}
\dtau=\frac{1}{\Ntrotter}.
\end{equation}
We construct the Suzuki-Trotter time-evolution operator $\UST$ in Eq.~\eqref{Eq:pMfMiteqone}  using the Dicke mapping and
evaluate the exact final population obtained without using the Suzuki-Trotter approximation,
\begin{equation}
p^{\text{exact}}_{\vec{\Ms}^{\rm f}\vec{\Ms}^{\rm i}}(t=\hbar/J)=
\left|\langle\vec{\Ms}^{\rm f} | 
e^{-i\Hamiltonian/J}
|\vec{\Ms}^{\rm i} \rangle \right|^2,
\end{equation}
and define the average discrepancy $\Avdisc$ as the difference between 
$p_{\vec{\Ms}^{\rm f}\vec{\Ms}^{\rm i}}$ and 
$p^{\text{exact}}_{\vec{\Ms}^{\rm f}\vec{\Ms}^{\rm i}}$ averaged over all the initial and final states,
\begin{equation}\label{Eq:DefDeltabar}
\Avdisc(t=\hbar/J) =\frac{1}{\zeta(\spin)}
\sum_{\vec{\Ms}^{\rm i},\vec{\Ms}^{\rm f}}
\left|p_{\vec{\Ms}^{\rm f}\vec{\Ms}^{\rm i}}(\hbar/J) -
p^{\text{exact}}_{\vec{\Ms}^{\rm f}\vec{\Ms}^{\rm i}}(\hbar/J)
\right|,
\end{equation}
where  $\zeta(\spin)=(2\spin+1)[1+2(2\spin+1)^2]/3$ is the total number of terms and only final states having the same $\Ms_{\text{tot}}$ as the initial state are included  in the sum over $\vec{\Ms}^{\rm f}$. The reason to use $\Avdisc$ defined in Eq.~\eqref{Eq:DefDeltabar} as the measure of the discrepancy instead of the wave function  fidelity or the matrix norm is that the expression \eqref{Eq:DefDeltabar} is straightforwardly evaluated  on a NISQ device or using a noise model.

\begin{figure}
	\includegraphics[width=\figwidth]{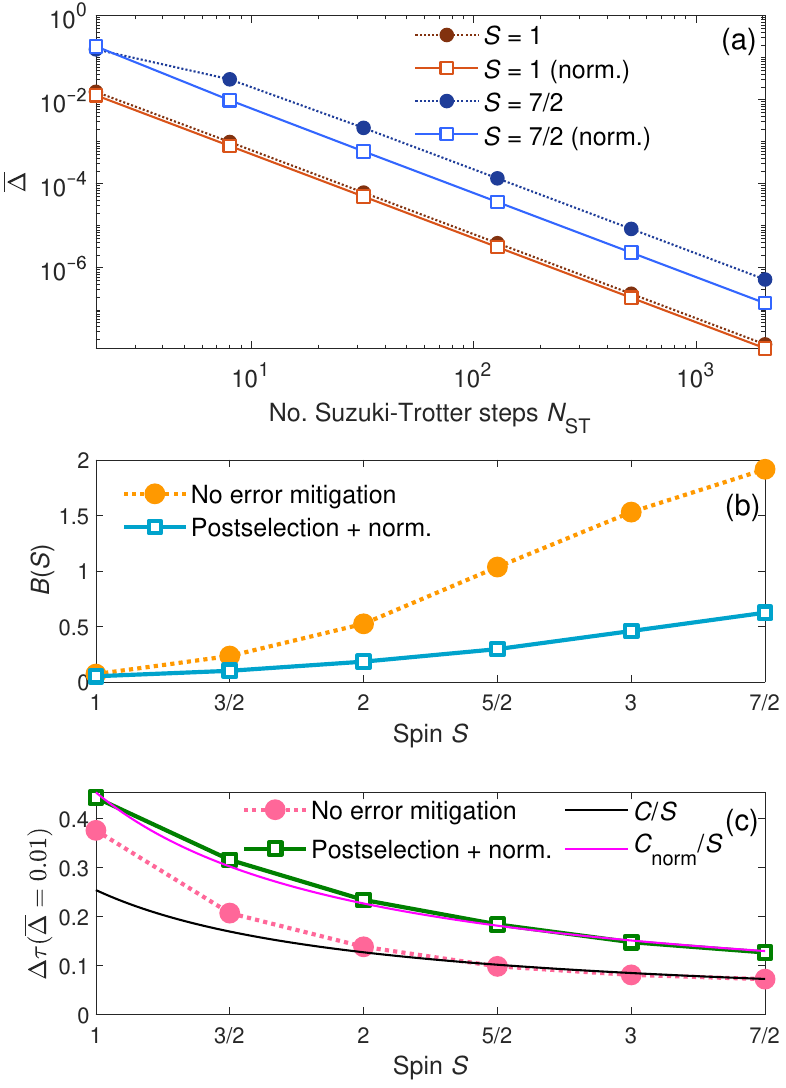}
	\caption{\label{Fig7}
		(a) The average difference $\Avdisc$ between the final spin-state population evaluated using the Suzuki-Trotter approximation in the absence of noise and the exact populations defined in Eq.~\eqref{Eq:DefDeltabar} evaluated at a fixed final time $t=\hbar/J$ as a function of the number of Suzuki-Trotter steps $\Ntrotter$. The curves labeled by ``norm.'' are calculated using populations whose error is mitigated by the postselection and normalization defined in Eq.~\eqref{Eq:mitigated_pnorm}.
		(b) Proportionality constant $B(\spin)$ in the fit $\Avdisc=B(\spin)/\Ntrotter^2$.
		(c) Time step size $\dtau$ required to achieve an average discrepancy $\Avdisc=0.01$. The filled circles and open squares are obtained  using the data in (b) and Eq.~\eqref{Eq:dtau_of_S} and the solid curves are best-fit curves obtained by the  least-squares fits to $\dtau=C/S$ with $C=0.254$ and $C_{\rm norm}=0.453$.
			}
\end{figure}

We first evaluate $\Avdisc(t=\hbar/J)$ in the ideal case of no noise. In Fig.~\ref{Fig7}(a), we show
$\Avdisc(t=\hbar/J)$ as a function of $\Ntrotter$ at  $\Ntrotter =2^x$ with $x=1,3,\ldots,11$ at two values of $\spin$, $\spin=1$ and $\spin=7/2$. Because of the Suzuki-Trotter approximation,  $\Ms_{\text{tot}}$ is not conserved in the Dicke mapping and, in general, 
$\sum_{\vec{\Ms}^{\rm f}} p_{\vec{\Ms}^{\rm f}\vec{\Ms}^{\rm i}}<0$ for the total final population. Note  that, in the noise-free case, the population leaks  not to spin states with $\Ms_{\rm tot}^{\rm f}\neq \Ms_{\rm tot}^{\rm i}$ but to qubit states not representing spin states. In other words, we 
have $p_{\vec{\Ms}^{\rm f}\vec{\Ms}^{\rm i}}=0$ for $\vec{\Ms}^{\rm f}\neq\vec{\Ms}^{\rm i}$. 
As discussed in Sec.~\ref{subsec:DickeMappingLargeS}, a simple form of error mitigation consists of the normalization of $p_{\vec{\Ms}^{\rm f}\vec{\Ms}^{\rm i}}$ to 1 given by
\begin{equation}\label{Eq:mitigated_pnorm}
 		p^{\rm norm}_{\vec{\Ms}^{\rm f}\vec{\Ms}^{\rm i}} =
 		\frac{p_{\vec{\Ms}^{\rm f}\vec{\Ms}^{\rm i}}}{\sum_{\vec{\Ms}^{' {\rm f}}}p_{\vec{\Ms}^{'{\rm f}}\vec{\Ms}^{\rm i}}}.
 \end{equation} 
The average discrepancy $\Avdisc$ evaluated using $p^{\rm norm}_{\vec{\Ms}^{\rm f}\vec{\Ms}^{\rm i}}$ in \eqref{Eq:DefDeltabar} is labeled by ``norm.'' in Fig.~\ref{Fig7}(a). As clearly seen in Fig.~\ref{Fig7}(a), 
the post-selected and renormalized $\Avdisc$ are smaller than $\Avdisc$ without error mitigation, demonstrating that the postselection can be used to mitigate the error in the Suzuki-Trotter approximation.      

We can also see in Fig.~\ref{Fig7}(a) that both the unmitigated $\Avdisc$ and the mitigated $\Avdisc$ are to a good approximation  proportional to 
$1/\Ntrotter^{2}$ [note that Fig.~\ref{Fig7}(a) is plotted in the double logarithmic scale]. We also find the same $1/\Ntrotter^{2}$ dependence of $\Avdisc$ for other values of $\spin<7/2$, which are not shown in Fig.~\ref{Fig7}(a).  The $1/\Ntrotter^{2}$ dependence of $\Avdisc$ may be surprising  because the Suzuki-Trotter approximation
\eqref{Eq:UST_qubit} is in general accurate to the first order in $\dtau$, that is, 
\begin{equation}
	\UST(\dtau)=e^{-i\dtau \sum_k h_kP_k} + \bigO(\dtau^2),
\end{equation}
and we would therefore expect the error in the wave function after $\Ntrotter$ steps to be proportional to $\dtau$ ($=1/\Ntrotter$). However,  in the Dicke mapping, we have
\begin{align}\label{Eq:DickeMappingOneTimeStep}
	e^{-i\dt \Hamiltonian_{\rm q}^{\rm Dicke}/\hbar}=&
	\prod_{k,l}e^{-i\dtau \W_{kl}}
	\nonumber\\
	&
	+\frac{\dtau^2}{4}\sum_{kl,k'l'}[\W_{kl},\W_{k'l'}]+\bigO(\dtau^3),
\end{align}
where $\W_{kl}$ was defined in Eq.~\eqref{Eq:DefWkl}, $[\cdot,\cdot]$ is the commutator, and the sum in the second line is taken over qubit indices $kl$ and $k'l'$ for which  either $k<k'$ or $l<l'$. Furthermore, we have 
\begin{equation}\label{Eq:UDickeWUdicke}
\braq{\SMone}
\UDicke^{\spin\dagger}
[\W_{kl},\W_{k'l'}]
\UDicke^{\spin}\ketq{\SMone}=0
\end{equation}
for the Dicke operator defined in Eq.~\eqref{Eq:UDicke_def}, because the commutator $[\W_{kl},\W_{k'l'}]$ is antisymmetric and the Dicke state $\ketq{D_{\spin,\Ms}}=\UDicke^{\spin}\ketq{\SMone}$ is symmetric with respect to the qubit indices $kl$ and $k'l'$. 
Equation \eqref{Eq:UDickeWUdicke} implies that the Suzuki-Trotter propagator $\prod_{k,l}e^{-i\dtau \W_{kl}}$ for one time step in the Dicke mapping is accurate to the second order in $\dtau$ because the term proportional to $\dtau^2$ in Eq.~\eqref{Eq:DickeMappingOneTimeStep} vanishes, and consequently, the error after $\Ntrotter$ steps becomes proportional to $\dtau^2=1/\Ntrotter^2$.

The average discrepancy $\Avdisc$ in Fig.~\ref{Fig7}(a) can be fitted with
\begin{equation}\label{Eq:Adisc_B}
	\Avdisc=\frac{B}{\Ntrotter^2},
\end{equation}
where $B$ is a fitting parameter. The resulting $\spin$-dependent $B(\spin)$ is shown in Fig.~\ref{Fig7}(b).
We can see that $B(\spin)$ increases as $\spin$ increases and that $B(\spin)$ obtained from the post-selected populations is smaller than the unmitigated $B(\spin)$ by a factor of two to three. Equation \eqref{Eq:Adisc_B} can be used to obtain the time step $\dtau$ required to achieve a certain average difference $\Avdisc$ at $t=\hbar/J$,
\begin{equation}\label{Eq:dtau_of_S}
	\dtau=\frac{1}{\Ntrotter}=\sqrt{\frac{\Avdisc}{B}}.
\end{equation}
The time step size $\dtau$ for an average discrepancy of $\Avdisc=0.01$ is shown in Fig.~\ref{Fig7}(c). We find that the value of $\dtau$ decreases  approximately as $\dtau=C/\spin$ for the large $\spin\ge2$. By fitting the data in Fig.~\ref{Fig7} for $S\ge 2$ to $\dtau=C/\spin$, we obtain 
$C=0.254$ and $C_{\rm norm}=0.453$.
Note that the numerical values shown in  Fig.~\ref{Fig7}(c) are obtained for $t=\hbar/J$. For the larger  final simulation time $T>\hbar/J$,  we expect that the discrepancy 
$\Avdisc$ becomes proportional to $T$ as $\Avdisc(T) = (JT/\hbar)\Avdisc(t=\hbar/J)$. Therefore, an estimate of $\dtau$ required to have an average discrepancy $\Avdisc_T$ at a final time $T>\hbar/J$ becomes
\begin{equation}\label{Eq:dtau_of_S2}
	\dtau=\sqrt{\frac{\Avdisc_T}{(JT/\hbar)B}},
\end{equation}
where $B(\spin)$ is the $\spin$-dependent proportionality constant derived for $T=\hbar/J$ [see Fig.~\ref{Fig7}(b)].

The case of $\spin=1/2$ is not considered in Fig.~\ref{Fig7} because the Suzuki-Trotter approximation is exact for a two-site, $\spin=1/2$ lattice. We have
\begin{align}
	e^{-it\Hamiltonian_{\rm q}^{\rm Dicke}/\hbar} &=
	e^{-i\frac{Jt}{4\hbar}(X_1X_0+Y_1Y_0+Z_1Z_0)}
	\nonumber\\
	&=
	e^{-i\frac{Jt}{4\hbar}X_1X_0}e^{-i\frac{Jt}{4\hbar}Y_1Y_0}e^{-i\frac{Jt}{4\hbar}Z_1Z_0}
\end{align} 
because the operators $X_1X_0$, $Y_1Y_0$, and $Z_1Z_0$ commute with each other.

\begin{figure}
	\includegraphics[width=\figwidth]{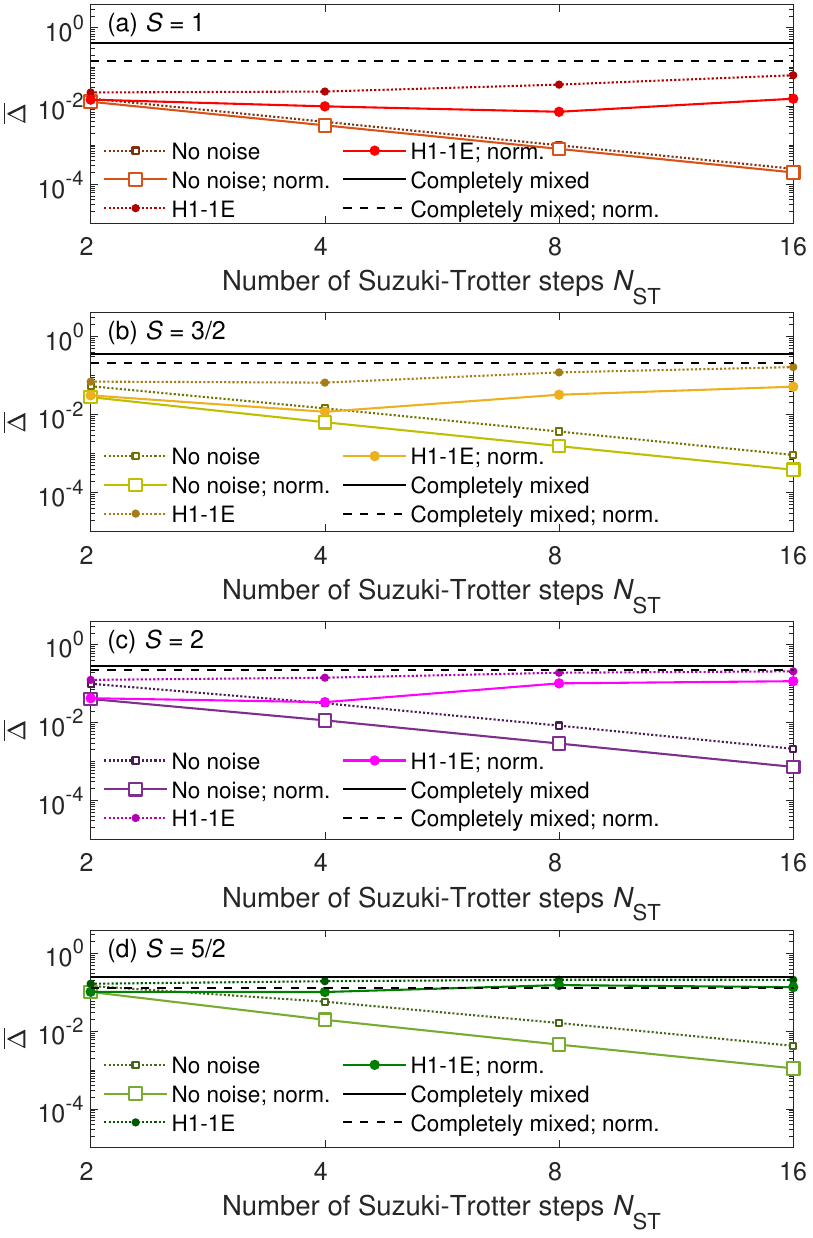}
	\caption{\label{Fig8}
	The average discrepancy $\Avdisc$ as defined in Eq.~\eqref{Eq:DefDeltabar} evaluated without noise (open squares) and with noise by the H1-1 emulator (filled circles) for (a) $\spin=1$, (b) $\spin=3/2$, 
	(c) $\spin=2$, and (d) $\spin=5/2$. The error bars for the H1-1E curves  estimated by bootstrapping are smaller than $0.01$ for all curves and are therefore not shown. The curves labeled by ``norm'' are the results obtained by the postselection and normalization according to Eq.~\eqref{Eq:mitigated_pnorm}. 
	The black horizontal lines represent the value of $\Avdisc$ obtained by assuming a completely mixed state in which all the qubit states are equally populated.}
\end{figure}

In order to investigate the effect of the noise on $\Avdisc$, we evaluate $\Avdisc(t=\hbar/J)$ as a function of $\Ntrotter$ for $1\le \spin \le 5/2$ using the H1-1 emulator \cite{QuantinuumH11E_2024}.  The results of the simulation in which $\Nshots=1024$ is chosen are shown in Fig.~\ref{Fig8}. For comparison, we also show the 
value of $\Avdisc$ obtained by assuming a completely mixed final state in which all qubit-states are equally populated, that is,  
\begin{equation}\label{Eq:completelymixedpop}
		p_{\vec{\Ms}^{\rm f}\vec{\Ms}^{\rm i}}(t=\hbar/J)=
		\frac{1}{2^{2\nqubit}}
\end{equation}
for all $\vec{\Ms}^{\rm f}$ and $\vec{\Ms}^{\rm i}$, where
$\nqubit=2\spin$ is the number of qubits per lattice site. The average discrepancy $\Avdisc_{\rm mix}$ obtained using 
Eq.~\eqref{Eq:completelymixedpop} can be regarded as the worst case where the effect of the noise is very large. A successful simulation should result in a value of $\Avdisc$ much smaller than $\Avdisc_{\rm mix}$.

In Fig.~\ref{Fig8}, we can see that  the noise-free curves (open squares) decrease with $\Ntrotter$ as $\Avdisc\propto 1/\Ntrotter^2$ for all the $\spin$ values. As can be seen in Figs.~\ref{Fig8}(a) ($\spin = 1$) and \ref{Fig8}(b) ($\spin = 3/2$),  the error-mitigated curves obtained using H1-1E decrease until an optimal value of $\Ntrotter=\Ntrotter^{\rm opt}$,  but $\Avdisc$ starts increasing when $\Ntrotter$ is further increased,  showing that the simulation becomes less accurate. 
For the larger $\spin$ values, the situation becomes worse as shown in Figs.~\ref{Fig8}(c) and ~\ref{Fig8}(d).

Because of the presence of the noise, there is a trade-off between the accuracy of the Suzuki-Trotter approximation which can be raised by increasing the number of Suzuki-Trotter steps $\Ntrotter$ and the effect of the noise which can also be raised by increasing $\Ntrotter$ associated with the increase in the depth of the quantum circuit. 
For the Dicke mapping applied to a two-site lattice the results shown in Fig.~\ref{Fig8} obtained using  the H1-1 emulator (two-qubit gate error $\errortwoq\approx 10^{-3}$)  suggest that the optimal number of Suzuki-Trotter steps is $\Ntrotter^{\rm opt}\approx 8$
for $\spin=1$ and $\Ntrotter\approx 4$ for $\spin=3/2$.
In the case of $\spin=2$, shown in Fig.~\ref{Fig8}(c), the average discrepancy for $\Ntrotter=2$, 
$\Avdisc_{\text{H1-1E}}^{\text{norm.}}(\Ntrotter =2)=0.103$, is almost the same as that for $\Ntrotter=4$, $\Avdisc_{\text{H1-1E}}^{\text{norm.}}(\Ntrotter =4)=0.102$, which means that the accuracy of the simulation is not improved by increasing the number of Suzuki-Trotter steps.
In Fig.~\ref{Fig8}(d), we see that the average discrepancy $\Avdisc$ obtained using H1-1 is close to the $\Avdisc_{\rm mix}$ obtained from the completely mixed state and conclude that $\spin=5/2$ is too large for accurate simulations on H1-1 at the present two-qubit gate error ($\errortwoq\approx 10^{-3}$).

\section{Summary}\label{Sec:Summary}
In the present study, we have compared four different  mappings of the Heisenberg model for a lattice of interacting spins with arbitrary $\spin$ to qubit and qudit forms. We have found that the Dicke mapping, in which the state of  a
spin-$\spin$ lattice site is described by a superposition of $2\spin$-qubit states, is the most efficient and accurate mapping, because the number of terms in the Hamiltonian is much smaller than that in the other qubit mapping schemes. The Dicke mapping has been assessed by simulating two-site and four-site lattices with $\spin$ up to $5/2$ using the Quantinuum trapped-ion quantum computer H1-1. Accurate time-dependent spin-state populations and expectation values have been obtained for $\spin\le 2$.
The Dicke mapping introduces couplings between non-neighboring qubits and is therefore particularly efficient for quantum computers having an all-to-all connectivity. 
The most  efficient large-$S$ encoding for quantum computers having a nearest-neighbor qubit connectivity requires a separate investigation.

We have found that a qudit mapping scheme, in which each lattice site is described by a qudit having $2\spin+1$ levels, results in an equally compact expression of the qudit Hamiltonian as the Dicke mapping. However, given that the gate errors of qudit devices are currently much larger than the gate errors of  qubit devices, the good performance of the Dicke mapping suggests that the Dicke mapping is more suited for the simulation of large $\spin$ lattices than a qudit mapping.

The  efficiency of the Dicke mapping demonstrated both theoretically and by simulations using current quantum hardware shows  that the Dicke mapping will be a powerful tool in the quantum computing of large-$\spin$ lattices  such as single-molecule magnets and molecular cluster complexes.

\acknowledgments
We thank K.\ Seki and S.\ Yunoki (RIKEN, Japan) for helpful comments.
This paper is based on results obtained in project JPNP20017, commissioned by the New Energy and Industrial Technology Development Organization (NEDO).
We are supported by the JSPS (Kakenhi no.~JP24K08336) and JST-CREST Quantum Frontiers (grant no.~JPMJCR23I7).  We thank the DIC Corporation for their support through the Applied Quantum Chemistry by Qubits
(AQUABIT) project under the UTokyo Quantum Initiative.

\appendix

\section{Number of terms in the compact mapping Hamiltonian}\label{Appendix_NtemrsHcompact}
Although  accurate simulations for very large values of $\spin>10$ are practically impossible, 
it is of theoretical interest to investigate the large-$\spin$ scaling of the number of terms $\NHq$ in the qubit Hamiltonian. In the case of the compact mapping, we cannot easily derive  an analytical formula of $\NHq$,  and therefore, we carry out  a numerical investigation for $1/2\le \spin \le 63/2$.

\begin{figure}
	\includegraphics[width=\figwidth]{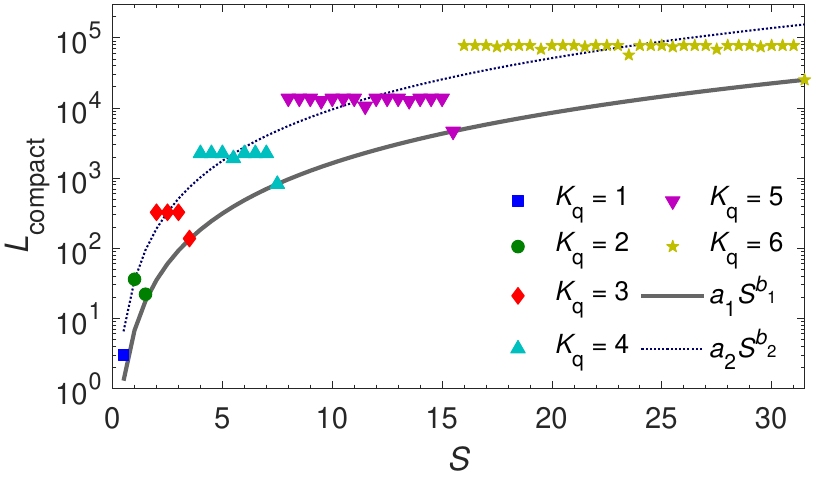}
	\caption{\label{Fig1_appendix}
		Number of terms $\NHq_{\rm compact}$	in the compact mapping  qubit Hamiltonian for a two-site lattice as a function of $\spin$.
		The data are grouped according to the required number of qubits per site $\nqubit$. The least-squares fit $a_1\spin^{b_1}$ with $a_1=6.7$ and $b_1=2.4$ is shown by a solid line, and $a_2\spin^{b_2}$ with $a_2=35.1$ and $b_2=2.4$ ($=b_1$) is shown by a dotted line.
	}
\end{figure}

The number of terms $\NHqcompact$ in the compact-mapping qubit Hamiltonian for a two-site lattice is shown in Fig.~\ref{Fig1_appendix} as a function of $\spin$. The number of qubits $\nqubit$ required per site is indicated by the color and marker style.  We can see that for values of $\spin$ requiring the same number of $\nqubit$, $\NHqcompact$ is roughly constant. The exception is the last data point in each group at $2^{\nqubit}=2\spin+1$, which means that all qubit states are used to represent spin states. In this case, $\NHqcompact$ is much smaller than the value obtained at the other values of $\spin$ for the same $\nqubit$.

Even though we have not been able to derive an analytical formula for $\NHqcompact(\spin)$, 
we find that $\NHqcompact$ in the range of $\spin$ shown in Fig.~\ref{Fig1_appendix} can be well fitted by a power law. For $\spin$ satisfying $2\spin+1=2^{\nqubit}$ ($\nqubit=1,\ldots,6$), we find that $\NHqcompact(\spin)\approx a_1\spin^{b_1}$ provides a good fit with $a_1=6.7$ and $b_1=2.4$. This curve is shown with a solid line in Fig.~\ref{Fig1_appendix}. In order to discuss the other values of $\spin$ where $\log_2(2\spin+1)$ are not an integer, we define the central value $\spin_{\rm c}$ for each $\nqubit$ as
\begin{equation}
	\spin_{\rm c}=
	\begin{cases}
		\frac{1}{2},& \text{if } \nqubit=1,\\
		3\times 2^{\nqubit -3}-\frac{1}{2},& \text{otherwise},
	\end{cases}
\end{equation}
and the average $\avNHqcompact(\nqubit)$ of $\NHqcompact(\spin)$ for $\spin$ corresponding to the same value of $\nqubit$, excluding $\spin=2^{\nqubit -1 }-1/2$ as
\begin{equation}
	\avNHqcompact(\nqubit)=
	\begin{cases}
		\NHqcompact(1),& \text{if } \nqubit =1,\\
		{\displaystyle\frac{1}{2^{\nqubit-1}}}{\displaystyle \sum_{\spin=2^{\nqubit-2}}^{2^{\nqubit-1}-1}}\NHqcompact(\spin),& \text{otherwise}.
	\end{cases} 
\end{equation}
We then fit $\avNHqcompact(\nqubit)$ by a power law of the form $a_2\spin_{\rm c}^{b_2}$ and find a good fit for $a_2=35.1$ and $b_2=2.4$. This fit is shown by a dotted line in Fig.~\ref{Fig1_appendix}. Note that the exponents for the solid and dotted curves in Fig.~\ref{Fig1_appendix} are the same, $b_1=b_2$, which implies that  both $\NHqcompact(\spin)$ with $\spin=2^{\nqubit-1}-1/2$ and $\NHqcompact(\spin)$ with $\spin\neq 2^{\nqubit-1}-1/2$ scale in the same manner when $\spin$ increases.

\section{Generalized Gell-Mann matrices}\label{Appendix_GellMann}
The generalized $\quditd\times\quditd$ Gell-Mann matrices $\GellMann_k$ ($k=1,2,\ldots,\quditd^2$), which 
provide a convenient  basis for $\quditd\times\quditd$ Hermitian matrices, are defined as follows \cite{Luo2014}. $\GellMann_1$ is defined as being proportional to the identity matrix,
\begin{equation}
	(\GellMann_1)_{mn}=\sqrt{\frac{2}{\quditd}} \delta_{mn},
\end{equation}
where $\delta_{mn}$ is the Kronecker delta.
The rest of the generalized Gell-Mann matrices are of three types: $X$-like, having matrix elements defined as 
\begin{equation}
	(\GellMann_k)_{mn} = \delta_{jm}\delta_{ln} +\delta_{jn}\delta_{lm},
\end{equation}
and $Y$-like,
\begin{equation}
	(\GellMann_k)_{mn} = -i\delta_{jm}\delta_{ln} +i\delta_{jn}\delta_{lm},
\end{equation}
for all $j$ and $l$ in the range
$ 1<j\le \quditd$, $1\le l <j$, and diagonal $Z$-like matrices defined for all $1<j\le d$ as 
\begin{equation}
	(\GellMann_k)_{mn} = \sqrt{\frac{2}{j(j-1)}}\Omega_m(j)\delta_{mn}
\end{equation}
with 
\begin{equation}
	\Omega_m(j)=\begin{cases}
		1,& \text{if } 1\leq m < j,\\
		1-j,& \text{if } m=j, \\
		0,& \text{if } m>j.
	\end{cases}
\end{equation}
We order the generalized Gell-Mann matrices so that $\GellMann_2$ corresponds to $j=2$, $l=1$ ($X$-like),
$\GellMann_3$ to $j=2$, $l=1$ ($Y$-like), $\GellMann_4$ to $j=2$ ($Z$-like), $\GellMann_5$ and $\GellMann_6$
to $j=3$, $l=1$ ($X$ and $Y$-like), $\GellMann_7$ and $\GellMann_8$
to $j=3$, $l=2$ ($X$ and $Y$-like), $\GellMann_9$
to $j=3$ ($Z$-like), and so on.

The generalized Gell-Mann matrices $\GellMann_k$ satisfy $\trace(\GellMann_k) =0$ and $\trace(\GellMann_k\GellMann_l) = 2\delta_{kl}$. For $\quditd=2$, they reduce to the standard Pauli matrices, and for $\quditd=3$, they reproduce the original Gell-Mann matrices \cite{GellMann1962}.

\begin{widetext}
\section{Qubit and qudit Hamiltonians for \boldmath{$\spin=1$}}\label{Appendix_quditHamiltoniansS1}
The qubit form of the  Hamiltonian for the compact mapping ($\spin=1$) is
\begin{align}
	\Hamiltonian_{\rm q}^{\rm compact}={}
	\frac{J}{8}(&
	2ZIZI+2ZIZZ+2ZZZI+2ZZZZ+IXIX
     +IXZX+IYIY+IYZY+ZXIX
     \nonumber
     \\
     &
     +ZXZX
     +ZYIY+ZYZY
     +IXXX+IXYY-IYXY
     +IYYX+ZXXX+ZXYY
     \nonumber
     \\
     &
     -ZYXY+ZYYX
     +XXIX+XXZX-XYIY-XYZY
     +YXIY
     +YXZY+YYIX
     \nonumber
     \\
     &
     +YYZX+XXXX+XXYY
     +XYXY-XYYX-YXXY+YXYX+YYXX
     +YYYY),
\end{align}

the qubit form of the direct Hamiltonian is 
\begin{align}
	\Hamiltonian_{\rm q}^{\rm direct}={}
	\frac{J}{8}(&2IIZIIZ+IXXIXX+IYYIYY+IXXIYY
     +IYYIXX-IYXIXY+IYXIYX
     \nonumber
     \\
     &
     +IXYIXY
     -IXYIYX-2IIZZII
     +IXXXXI+IYYYYI
     +IXXYYI+IYYXXI
     \nonumber
     \\
     &
     -IYXXYI+IYXYXI
     +IXYXYI-IXYYXI+XXIIXX
     +YYIIYY
     +XXIIYY
     \nonumber
     \\
     &
     +YYIIXX-YXIIXY+YXIIYX
     +XYIIXY-XYIIYX+XXIXXI+YYIYYI
     \nonumber
     \\
     &
     +XXIYYI+YYIXXI-YXIXYI+YXIYXI
     +XYIXYI-XYIYXI-2ZIIIIZ
          \nonumber
     \\
     &
     +2ZIIZII),
\end{align}
the qubit form of the Dicke Hamiltonian is 
\begin{equation}
	\Hamiltonian_{\rm q}^{\rm Dicke}=
	\frac{J}{4} (XIXI+YIYI+ZIZI+XIIX+YIIY+ZIIZ
     +IXXI+IYYI+IZZI+IXIX+IYIY+IZIZ),
     \end{equation}
     and the qudit ($d=3$) Hamiltonian is
     \begin{equation}
	\Hamiltonian_{d}=
	\frac{J}{2} (
     \lambda_2\lambda_2+\lambda_2\lambda_7+\lambda_3\lambda_3+\lambda_3\lambda_8+
     \frac{1}{2}\lambda_4\lambda_4+a\lambda_4\lambda_9
     +\lambda_7\lambda_2+\lambda_7\lambda_7+\lambda_8\lambda_3+\lambda_8\lambda_8+a\lambda_9\lambda_4+\frac{3}{2}\lambda_9\lambda_9),
     \end{equation}
     where $a\approx 0.8660$, and  $\GellMann_k$ is a generalized Gell-Mann matrix
     defined in Appendix \ref{Appendix_GellMann}.
  
\section{Qubit and qudit Hamiltonians for \boldmath{$\spin=3/2$}}\label{Appendix_quditHamiltoniansS3half}
The qubit form of the  Hamiltonian for the compact mapping ($\spin=3/2$) is
\begin{align}
	\Hamiltonian_{\rm q}^{\rm compact}={}
	\frac{J}{4}(&IZIZ+2IZZI+2ZIIZ+4ZIZI+3IXIX
     +3IYIY+cIXXX+cIXYY
     \nonumber
     \\
     &
     -cIYXY+cIYYX+cXXIX
     -cXYIY+cYXIY+cYYIX+XXXX
     +XXYY
     \nonumber
     \\
     &
     +XYXY-XYYX-YXXY+YXYX
     +YYXX
     +YYYY),
\end{align}     
     where $c\approx 1.7321$.
     The qubit form of the  direct mapping Hamiltonian  is
\begin{align}
	\Hamiltonian_{\rm q}^{\rm direct}={}
	\frac{3J}{16}(& 3IIIZIIIZ+IIIZIIZI+IIXXIIXX
     +IIYYIIYY+IIXXIIYY+IIYYIIXX
     \nonumber
     \\
     &
     -IIYXIIXY+IIYXIIYX
     +IIXYIIXY
     -IIXYIIYX-IIIZIZII
     \nonumber
     \\
     &
     +fIIXXIXXI+fIIYYIYYI+fIIXXIYYI+fIIYYIXXI
     -fIIYXIXYI
     \nonumber
     \\
     &
        +fIIYXIYXI+fIIXYIXYI-fIIXYIYXI-3IIIZZIII+IIXXXXII
     \nonumber
     \\
     &
     +IIYYYYII
     +IIXXYYII+IIYYXXII-IIYXXYII+IIYXYXII+IIXYXYII
     \nonumber
     \\
     &
     -IIXYYXII+IIZIIIIZ+\frac{1}{3}IIZIIIZI
     +fIXXIIIXX+fIYYIIIYY     
     \nonumber
     \\
     &
     +fIXXIIIYY
     +fIYYIIIXX-fIYXIIIXY+fIYXIIIYX
     +fIXYIIIXY
     \nonumber
     \\
     &
     -fIXYIIIYX
     -\frac{1}{3}IIZIIZII
     +\frac{4}{3}IXXIIXXI+\frac{4}{3}IYYIIYYI+\frac{4}{3}IXXIIYYI
     \nonumber
     \\
     &
     +\frac{4}{3}IYYIIXXI
     -\frac{4}{3}IYXIIXYI
     +\frac{4}{3}IYXIIYXI
     +\frac{4}{3}IXYIIXYI-\frac{4}{3}IXYIIYXI
     \nonumber
     \\
     &
     -IIZIZIII
     +fIXXIXXII+fIYYIYYII+fIXXIYYII
     +fIYYIXXII
     \nonumber
     \\
     &
     -fIYXIXYII
     +fIYXIYXII
     +fIXYIXYII-fIXYIYXII-IZIIIIIZ
     \nonumber
     \\
     &
     -\frac{1}{3}IZIIIIZI
     +XXIIIIXX+YYIIIIYY
     +XXIIIIYY+YYIIIIXX-YXIIIIXY
     \nonumber
     \\
     &
     +YXIIIIYX+XYIIIIXY-XYIIIIYX
     +\frac{1}{3}IZIIIZII+fXXIIIXXI
     \nonumber
     \\
     &
     +fYYIIIYYI
     +fXXIIIYYI+fYYIIIXXI-fYXIIIXYI
     +fYXIIIYXI
     \nonumber
     \\
     &
     +fXYIIIXYI-fXYIIIYXI
     +IZIIZIII+XXIIXXII+YYIIYYII
     \nonumber
     \\
     &
     +XXIIYYII+YYIIXXII-YXIIXYII
     +YXIIYXII+XYIIXYII
     \nonumber
     \\
     &
     -XYIIYXII
     -3ZIIIIIIZ-ZIIIIIZI+ZIIIIZII
     +3ZIIIZIII,
     \end{align}
     where $f \approx 1.1547$. 
     
     The qubit form of the  Dicke Hamiltonian  is
\begin{align}
	\Hamiltonian_{\rm q}^{\rm Dicke}={}
	\frac{J}{4}(& XIIXII+YIIYII+ZIIZII+XIIIXI
     +YIIIYI+ZIIIZI+XIIIIX+YIIIIY
          \nonumber
     \\
     &
+ZIIIIZ+IXIXII+IYIYII
     +IZIZII
     +IXIIXI+IYIIYI+IZIIZI+IXIIIX
          \nonumber
     \\
     &
+IYIIIY+IZIIIZ+IIXXII+IIYYII
     +IIZZII+IIXIXI
     +IIYIYI+IIZIZI
     \nonumber
     \\
     &
     +IIXIIX+IIYIIY+IIZIIZ.
     \end{align}
     
     The qudit Hamiltonian for $\spin=3/2$ ($d=4$), defined in terms of the generalized Gell-Mann matrices $\GellMann_k$, is
     \begin{align}
     \Hamiltonian_{d}={}
	\frac{3J}{4}(& 
     \lambda_2\lambda_2+b_1\lambda_2\lambda_7+\lambda_2\lambda_{14}+\lambda_3\lambda_3+b_1\lambda_3\lambda_8+\lambda_3\lambda_{15}
     +\frac{1}{3}\lambda_4\lambda_4+b_2\lambda_4\lambda_9
     +b_3\lambda_4\lambda_{16}
     \nonumber
     \\
     &
     +b_1\lambda_7\lambda_2+\frac{4}{3}\lambda_7\lambda_7
     +b_1\lambda_7\lambda_{14} 
     +b_1\lambda_8\lambda_3+\frac{4}{3}\lambda_8\lambda_8+b_1\lambda_8\lambda_{15}+b_2\lambda_9\lambda_4
     +\lambda_9\lambda_9
     \nonumber
     \\
     &
     +\sqrt{2}\lambda_9\lambda_{16}
     +\lambda_{14}\lambda_2+b_1\lambda_{14}\lambda_7+\lambda_{14}\lambda_{14}
     +\lambda_{15}\lambda_3+b_1\lambda_{15}\lambda_8
     +\lambda_{15}\lambda_{15}
     +b_3\lambda_{16}\lambda_4
\nonumber
     \\
     &
          +\sqrt{2}\lambda_{16}\lambda_9+2\lambda_{16}\lambda_{16}),
     \end{align}
     where $b_1\approx 1.1547$, $b_2\approx 0.57735$, and $b_3\approx 0.8165$.

     \end{widetext}

%

\end{document}